\documentclass[onecolumn,fleqn,usenatbib]{mnras}
\usepackage[T1]{fontenc}
\usepackage{ae,aecompl}
\usepackage{amsfonts}
\usepackage{amssymb}
\usepackage{amsmath}
\usepackage{graphicx}
\usepackage{epstopdf}
\usepackage{float}
\usepackage{soul}
\newcommand{\vel}{\boldsymbol{{u}}}

\title{Convective overshooting and penetration in a Boussinesq spherical shell}

\author[L. Korre et al.]{L. Korre,
$^{1}$\thanks{E-mail:lkorre@ucsc.edu}
P. Garaud,$^{1}$and 
N. Brummell$^{1}$
\\
$^{1}$
Department of Applied Mathematics and Statistics, Jack Baskin School of Engineering, \\
University of California Santa Cruz,
1156 High Street, Santa Cruz, California 95064, USA
}
\date{Accepted XXX. Received YYY; in original form ZZZ}

\pubyear{2018}

\begin{document}
\label{firstpage}
\pagerange{\pageref{firstpage}--\pageref{lastpage}}
\maketitle

\begin{abstract}
We study the dynamics associated with the extension of turbulent convective motions from a convection zone (CZ) into a stable region (RZ) that lies below the latter. For that purpose, we have run a series of three-dimensional direct numerical simulations  solving the Navier-Stokes equations under the Boussinesq approximation in a spherical shell geometry. We observe that the overshooting of the turbulent motions into the stably stratified region depends on three different parameters: the relative stability of the RZ, the transition width between the two, and the intensity of the turbulence. In the cases studied, these motions manage to partially alter the thermal stratification and induce thermal mixing, but not so efficiently as to extend the nominal CZ further down into the stable region. We  find that the kinetic energy below the convection zone can be modeled by a half-Gaussian profile whose amplitude and width can be predicted a priori for all of our simulations. We examine different dynamical lengthscales related to the depth of the extension of the motions into the RZ, and we find that they all scale remarkably well with a lengthscale that stems from a simple energetic argument. We discuss the implications of our findings for 1D stellar evolution calculations. 

\end{abstract}

\section{Introduction}
\label{sec:intro}
Understanding the dynamical interaction between an unstable turbulent convective region and an adjacent stable one remains a long-standing unsolved problem in fluid dynamics. This situation is nevertheless fairly ubiquitous in both geophysical and astrophysical settings. Here on Earth for example, it commonly occurs in the atmosphere, where re-radiance of solar surface warming creates a mixed layer below the very stable nocturnal inversion layer \citep{Deardorff69}. In stars, which are the main topic of this paper, the coexistence of convective and radiative layers is almost ubiquitous across masses and evolutionary stages. For example, A-type stars possess two convection zones, an upper one driven predominantly by the ionization of hydrogen and a lower one driven by the second ionization of helium, with a radiative zone in between.  In the Sun, by contrast, an outer convection zone sits atop an inner radiative zone and below a stable atmosphere, and the transitions are due to changes in the heat capacity (caused by the partial ionization zones of hydrogen and helium) and opacity (due to the temperature dependence of heavier ions).  Since there is no impermeable interface between the stable and unstable layers, fluid flows originating from one can continue into the other.  Primary questions are then whether the convective region can be extended from its original size, and whether the stability characteristics of the system are altered significantly. In all that follows, we shall adopt the terminology introduced by \citet{Zahn91}: if motions are found beyond the convective layer but do not extend its size then the dynamics are termed ``overshooting"; if the convective region is extended, the dynamics are called ``penetration".  

Any form of mixing beyond the classical boundary set by the Schwarzschild criterion could have crucial impacts on stellar evolution and surface abundances through the transport of chemicals and angular momentum \citep[e.g.][]{Straus76, Spite82,Ahrens92, Pinsonneault, Herwig2000, Baraffe17}. The transport of magnetic fields between the two regions has also been suggested as playing a major role in the dynamo process \citep[e.g.][]{vanBallegooijen, Parker93,Charbonneau97}. Furthermore, thanks to the development of helio- and asteroseismology, we now have the opportunity to directly measure the extent of an adiabatically stratified  zone \citep[e.g.][]{Christensen-Dalsgaard91, SilvaAguirre2011}. This provides a direct test of stellar evolution, and can in particular reveal the presence of penetration beyond the expected edge of a convective region \citep[e.g.][]{Deheuvels2016, Christensen-Dalsgaard2018}. Because of its obvious importance, a great body of work has already been generated on penetrative and overshooting convection, and yet some of the crucial questions remain poorly understood. In what follows, we summarize some of the salient modeling milestones of the field, review any outstanding questions and place our work in their context.  

The answer to the most basic question of ``does penetration or overshooting actually happen?" has been addressed using a classic example of such dynamics, the ice-water system. Adjacent convective and stable regions in this system can be created thanks to the unusual equation of state for water, which is quadratic with a density maximum at $4^{\circ}$ C.  When a layer of water sits on top of ice (at $0^{\circ}$ C) with an upper boundary temperature of more than $4^{\circ}$ C, a system is created where a convectively-unstable layer (between the ice and the location of the density maximum) sits below a convectively-stable layer (above the density maximum).   A long history of exploration of this problem exists, from experiments \citep[e.g.][]{Malkus, Townsend}, through linear and weakly nonlinear analytic work \citep[e.g.][]{Veronis63}, to numerical simulations both old \citep[e.g.][]{Musman68, MW73} and new \citep[e.g.][]{Couston17}.   This simple toy model clearly demonstrates that the weak overshooting predicted by linear theory is replaced by deeper penetration when nonlinear feedback on the thermal stratification is allowed. This raises the crucial question of whether similarly large deviations from linear theory predictions (the Schwarzschild criterion) exist in stars.

Answering this question requires moving beyond the assumptions of the works cited above, which was almost all two-dimensional and (essentially) incompressible.  A first attempt to understand three-dimensional and compressible effects numerically came from modeling via modal expansions, an approach that is motivated by the observable cellular nature of convection. The horizontal structure of the flow is expanded as a low-order discrete spectrum of horizontal planform modes, allowing numerical resources to be devoted to solving the vertical and temporal problem.  This approach, first introduced by \citet{Herring63} and \citet{Roberts66} but popularized in a series of papers by Gough, Spiegel and Toomre \citep{GST75, TGS77,TGS82}, was first applied to the B\'{e}nard convection problem, and commonly uses severe truncations of the modal expansion (1-3 modes) with planforms such as rolls, squares and hexagons.  

The technique has been extended to penetrative and overshooting problems involving multiple layers in a number of ways.  Using a complex equation of state to include the ionization regions, as well as the anelastic formalism, the papers by Latour et al. \citep{LSTZ76, TZLS76, LTZ81} study the convection zones of A-type stars, and find that they could interact despite the intervening radiative zone thanks to extended fluid motions.  
Somewhat later, \citet{ZTL82} and \citet{MLTZ84} simplified the model setup to address the question of penetration specifically. Using Boussinesq and anelastic equations respectively, they initiate layers by directly specifying a depth-dependent background adiabatic gradient, in order to study a single convection zone sandwiched between two stable layers. This compact series of papers has led to some important realizations.   Firstly, a fairly deep penetration on the order of the depth of the unstable layer is found in all cases which agrees well with laboratory experiments (although it depends on the stability of the stable layer and on the aspect ratio of the cells).  Secondly, flow asymmetries make substantial differences in the amount of overshoot or penetration. For instance, stratification combined with pressure effects (buoyancy braking in the upflows and enhanced driving in the downflows) in the anelastic case causes slower upflows and faster downflows compared to the Boussinesq case, which leads to enhanced downward penetration. Topological asymmetries (as induced by non-Cartesian effects or simply through a particular selection of horizontal planform) have similarly important impact on the problem. 

The discovery of the importance of flow asymmetries  on the extent of overshooting and penetration has naturally prompted new investigations into the effect of compressibility.  A big step forward was made by \citet{Hurlburt86,Hurlburt94}, with fully nonlinear, compressible, two-dimensional, Cartesian simulations of overshooting/penetrative convection. Note that these two papers also introduce yet another way of creating a radiative/convective system, by using a vertically-varying thermal conductivity profile. This creates a variation in the background radiative temperature gradient in the different layers which can be selected to achieve different stability properties, and the background model ultimately takes the form of stacked polytropes. This setup naturally introduces the concept of ``stiffness" $S$ as the ratio of the subadiabaticity of the stable region to the superadiabaticity of the unstable region. Most notably, these papers investigate the dependence of the depth $\delta$ of extended motions on the stiffness $S$, revealing two separate regimes: one associated with penetration ($\delta \propto S^{-1}$) and one with overshoot ($\delta \propto S^{-1/4}$) (more on this topic later). These studies also demonstrate the generation of gravity waves in the stable interior by the overshooting, both in fully compressible simulations (e.g. \citet{Hurlburt86}, see also \citet{Pratt17}) and in anelastic ones \citep[e.g.][]{Rogers2005,BMT11}. 
 \citet{Freytag96} performed fully compressible two-dimensional, radiation-hydrodynamics simulations of the narrow convection zones sandwiched between stable layers created by a complex equation of state including ionization found in A-type stars and cool DA white dwarfs. This paper notably finds deep overshooting, attributes the exponential drop off observed in the overshoot velocity to the stable ``tail" of the convectively unstable modes excited in the convection zone, and derives a depth-dependent diffusion coefficient to describe the corresponding mixing. This exponential formulation for mixing by overshooting convection is now commonly used in stellar evolution codes \citep[e.g.][]{Herwig2000,Paxton2011,Paxton2013,Sukhbold2014}.

Three-dimensional simulations of the Cartesian stacked polytropic model became possible in the latter part of the 1990s. \citet{Singh95}, \citet{Singh98} and \citet{Saikia2000} for instance present a series of low resolution large-eddy-simulations (LES) with sub-grid-scale (SGS) modelling while \citet{Muthsam95} present low resolution finite-difference models. All of these are fully compressible, and mostly appear to confirm the ideas of the two-dimensional simulations and analysis, including the various aforementioned scalings with the stiffness, $S$.  Somewhat later, however, \citet{Brummell} presented a more comprehensive parameter survey performed with high resolution, direct numerical simulations (DNS), including much more turbulent cases and a wider range of $S$. That work finds only overshooting and no true penetration, even in the parameter regimes where it would be most likely to occur, such as high Rayleigh number, low Prandtl number and low $S$. Instead, the transition from adiabatic to subadiabatic stratifications is seen to be rather smooth, and takes place across an extended partially mixed region. The authors attribute this mainly to the low filling factor of the downflowing convective plumes in the turbulent compressible case, arguing that the earlier  low resolution 3D models only found penetration because they were far more laminar and almost two-dimensional. 

In parallel with the predominantly numerical efforts described above, a variety of more phenomenological models have been  proposed to date. Early works in stellar evolution typically use a non-local formulation of mixing-length theory  \citep[e.g.][]{Spiegel63,ShavivSalpeter1973, Cogan75, Maeder75}, with results that vary widely depending on specific assumptions associated with the nonlocal integration scale, as criticized by \citet{Renzini}. As the aforementioned numerical simulations began to provide more insight into the dynamics of overshooting convection, phenomenological models have shifted towards a more realistic representations of the convective flows.   In addition to the semi-analytical weakly nonlinear theories discussed earlier, \citet{vanBallegooijen} for instance includes the effect of the horizontal flows near the base of the convection zone via linear convective roll modes with an assumed nonlinear saturation amplitude, with similar results to those of the mixing length theory (that only includes vertical motions).  \citet{Schmitt84} builds on the emerging idea that the convective motions are more plume-dominated than cellular by using a meteorological model for plumes in a stable stratification with entrainment \citep{Morton56}; the model formally reduces to the mixing length model of \citet{ShavivSalpeter1973} in the limit of zero entrainment. \citet{Schmitt84} find that shallow penetration is likely in the solar case, with the depth being dependent mainly on the velocity and filling factor of the plumes at the base of the convection zone (and insensitive to other parameters, such as the entrainment rate), and that the transition to radiative dynamics below likely takes place through a very thin  thermal adjustment boundary layer. 

The work of \citet{Zahn91} simplifies these ideas by applying scaling arguments to the problem. He separates the dynamics below the convection zone into a true penetrative region (where the motions are vigorous enough to mix the background stratification to adiabatic) and a thermal boundary layer containing overshooting. His model recovers the dependence of the penetration depth on the typical convection zone velocity ($\propto w^{3/2}$) and on the assumed filling factor  of the plumes ($\propto f^{1/2}$) found numerically by \citet{Schmitt84}, which is interesting since both models make rather different assumptions on the nature of the plumes. \citet{Zahn91} also finds that the depth of this layer depends on the gradient of the conductivity profile, leading to a value of about 50\% of a pressure scale height in the solar case. Finally, Zahn's thermal boundary layer is very thin as in \citet{Schmitt84}.  A similar model is used in \citet{Hurlburt94}, but with the smooth conductivity profile replaced by a more abrupt piecewise-constant one corresponding to their stacked polytrope numerical simulations.  Writing their predictions for the depth $\delta$ of the mixed layers in terms of the stratification (stiffness) ratio, $S$, they can explain their aforementioned observed scaling laws, namely  $\delta \propto S^{-1}$ for  true penetration and $\delta \propto S^{-1/4}$ for the thermal boundary (overshoot) layer. They explain the transition in the scalings with increasing $S$ as a tradeoff between the increase in buoyancy braking and the decrease in local P\'{e}clet number .   

\citet{Rempel2004} builds upon these previous works with a semi-analytical model that follows a distribution of plumes throughout both the convection zone and overshoot region and  includes their interaction with the upflows.  This model thereby essentially incorporates nonlocality and entrainment, and further allows departures from the parameter regimes where mixing length theory is most likely to work (i.e. towards parameter regimes accessible by numerical simulations). Its predictions mirror the findings of \citet{Schmitt84} and \citet{Zahn91} but also reveal the extra dependencies of the overshoot characteristics on the total energy flux (determining the vigor of the eddies in the convection zone) and the assumed degree of mixing by  entrainment. In particular, the dependence on the nonlocal convective efficiency is postulated to explain the presence of true penetration in mixing length results (which are necessarily highly turbulent) by contrast with its absence in the three-dimensional simulations (where the degree of turbulence is limited due to numerical issues). Furthermore, this approach demonstrates that an ensemble of plumes with a distribution of velocities behaves quite differently from one where all the plumes have the same assumed velocity. In particular, the former results in a much smoother thermal transition between the penetration layer and the deeper radiative stratification than the latter, which has important observational implications for helioseismology \citep[e.g.][]{Monteiro1994,Monteiro1998}. 

To summarize, the main robust conclusions of these numerical and phenomenological modeling efforts are that penetration and overshooting can take place down to some fraction of the pressure scale height that depends on the exit velocity and the filling factor (or scale) of the downflowing motions at the base of the convection zone. The velocity of dowflowing plumes depends on the  strength of the convection itself in a non-local, bulk sense, requiring high P\'{e}clet number for any chance of penetration. Meanwhile the filling factor of these plumes depends on many factors such as geometry (2D vs 3D), compressibility, stratification, and on a turbulent entrainment efficiency that remains poorly constrained. These models also reveal dynamical differences between smooth and abrupt transitions in the background stratification associated with both radiative and adiabatic temperature gradients \citep[e.g.][]{Zahn91,Rogers2005}. 

Moving forward, the next natural step towards a better understanding of overshooting and penetration should involve three-dimensional simulations in a spherical geometry and some effects of compressibility -- either using fully compressible equations or anelastic equations.  Although quite a number of simulations of this variety have actually been performed, the vast majority of them have not explicitly examined the penetrative/overshooting question, since they were directed at the global dynamo problem or the solar tachocline problem \citep[for recent efforts, see e.g.][]{Brun2005,Miesch2006,Browning2006, Browning2007,Ghizaru2010,Racine2011,BMT11,Strugarek2011}.  Since such dynamo-directed models require significant turbulence, the considerable expense of these computational efforts has been dedicated to a small number of  simulations that are the most relevant, rather than an exhaustive study of parameter space.  Notable exceptions are the work of \citet{Browning2004} and \citet{Brun2017}, who look at differential rotation and overshoot in core-convective stars and solar-type stars respectively. In both cases, however, the set of simulations presented are far from actual stellar parameters in terms of actual diffusivities, and vary quantities such as the rotation rate and/or the stellar mass, rather than input parameters that more directly control the strength of the convection and  the stratification of the nearby radiative zone. Because of this, the results cannot easily be used to form a prognostic model for overshoot and penetration in more stellar-like conditions.

This paper therefore presents a parametric survey of stellar-like overshooting convection in a three-dimensional spherical geometry using direct numerical simulations. As a first step towards understanding the full problem, we consider Boussinesq dynamics \citep{SV} only, arguing that in many instances the interface between radiative and convective regions is located very deep in the interior of the star where this approximation is not perfect but reasonable.  For example, the pressure scale height at the bottom of the solar CZ is  approximately  $1/3$ of the depth of the solar CZ. We also ignore rotation in order to isolate the effects of geometry (asymmetry) and of the model parameters.   Our goal is to quantitatively characterize various aspects of the dynamics of the overshoot/penetration zone, in particular, the relationship between the  typical velocity of convective eddies and the amount of mixing induced beyond the edge of the original convection zone. Ultimately, we shall answer the question of when one should expect overshoot or true penetration in a star, and provide a usable prescription for mixing by overshooting convection that can easily be incorporated into one-dimensional stellar structure models.

\par The paper is organized as follows. In Section \ref{sec:model}, we describe the model configuration along with the initial conditions and the boundary conditions. In Section \ref{sec:samplesim}, we present some general characteristics of  a canonical simulation and we describe three characteristic lengthscales. In Section \ref{sec:kinprof}, we provide a model for the kinetic energy profile below the base of the convection zone. In Section \ref{sec:thermal}, we focus on thermal mixing in the radiative zone due to the overshooting of the turbulent motions in the stable region.   Finally, in Section \ref{sec:ccl}, we summarize our results, provide  comparisons with previous numerical work, and discuss the implications of these results in the solar and stellar overshooting dynamics.

\section{MODEL SETUP}
\label{sec:model}

We are interested in studying a two-layered system, consisting of a convectively unstable zone (CZ) overlying a radiative zone (RZ) which is everywhere locally stable to convection according to the Schwarzschild criterion.   The numerical model used builds upon the purely convective spherical shell setup described in \cite{Korre}, extended to include a convectively stable inner spherical shell beneath the unstable one. Our chosen  shell has an outer radius $r_o$, and  inner radius $r_i = 0.2 r_o$, with the CZ-RZ interface located at $r_t = 0.7 r_o$. This geometry was chosen to mimic that of the Sun, as an example of a fairly typical low-mass star. The position of the inner boundary does not affect any of our results, as long as $r_i \ll r_t$. The selection of the convection zone aspect ratio $r_t / r_o$ is expected to affect the results, on the other hand. However, we have chosen to keep it fixed since there are already many other parameters that need to be varied in the simulations (see below).

In an attempt to be relevant for stellar contexts, we adopt a number of specific dynamical ingredients. We  solve the three-dimensional (3D) Navier-Stokes equations under the \citet{SV}  Boussinesq approximation, which takes into account a non-zero adiabatic temperature gradient to account for weak compressibility. We assume constant thermal expansion coefficient $\alpha$, viscosity $\nu$, thermal diffusivity $\kappa$, adiabatic temperature gradient $dT_{\rm{ad}}/dr$ and gravity $g$. Note that these quantities would of course not be constant over the range $r = [0.2r_o,r_o]$ in a star -- this assumption is made for simplicity.  We fix the heat flux at the inner boundary, to account for the energy generated from nuclear burning in the stellar core, whereas at the outer boundary we fix the temperature. While the latter does not realistically capture the more complex radiative transfer processes that are known to control the photospheric boundary conditions in solar-type stars, we use this approximation because it is simple, with the expectation that it does not affect the convective dynamics near the bottom of the convection zone. Finally, we perform all of our simulations in a low Prandtl number regime (where the Prandtl number is the ratio of the viscosity to thermal diffusivity), which is again more relevant in the astrophysical context. To the authors' knowledge, this is the first time that penetrative convection is being studied in a  Boussinesq spherical shell geometry with the temperature boundary conditions as described above, and in the low Prandtl number regime.

We  let $T(r,\theta,\phi,t)=T_{\rm{rad}}(r)+\Theta(r,\theta,\phi,t)$ where $T_{\rm{rad}}$ is the temperature profile our system would have under pure radiative equilibrium, and where $\Theta$ describes temperature fluctuations away from that radiative equilibrium. 
 As part of the Boussinesq approximation, a linear relationship is assumed between the temperature and density perturbations such that $\rho/\rho_m = - \alpha \Theta$, where $\rho_m$ is the mean density of the background fluid. Then, the governing Boussinesq equations  are \citep{SV}:  
\begin{equation}
\nabla\cdot \vel=0,
\end{equation}
\begin{equation}
\displaystyle\frac{\partial\vel}{\partial t}+\vel\cdot\nabla\vel=-\frac{1}{\rho_m}\nabla p+\alpha \Theta g\boldsymbol{e_r}+\nu\nabla^2\vel,
\end{equation}
and
\begin{equation}
\displaystyle\frac{\partial\Theta}{\partial t}+\vel\cdot\nabla\Theta+u_r\left(\frac{dT_{\rm{rad}}}{dr}-\frac{dT_{\rm{ad}}}{dr}\right)=\kappa\nabla^2\Theta,
\end{equation}
where $\vel=(u_r,u_{\theta},u_{\phi})$ is the velocity field and $p$ is the pressure  perturbation away from hydrostatic equilibrium.

One way to set up the desired two-layered system is by ensuring that 
$dT_{\rm{rad}}/dr - dT_{\rm{ad}}/dr$ is negative in the CZ, and positive in the RZ. Since $T_{\rm{rad}}$ is the  temperature profile at  radiative equilibrium, and since we considered that $\nu$ and $\kappa$ are constant, the only way to ensure that its gradient changes significantly (aside from geometric effects) is to assume the existence of a heating source  localized near $r_t$, such that
\begin{equation}
\label{eq:HS}
\kappa\nabla^2 T_{\rm{rad}}=-H_s(r).
\end{equation}
The function $T_{\rm{rad}}(r)$ is the solution of this equation, with the boundary conditions

\begin{equation}
-\kappa \frac{dT_{\rm rad}}{dr} \bigg|_{r=r_i}  = F_{\rm{rad}},  
\end{equation} 
where $F_{\rm{rad}}$ is the temperature flux per unit area through the inner boundary, and 
\begin{equation}
T(r_o) = T_o.
\end{equation} 
Integrating Equation (\ref{eq:HS}) once yields

\begin{equation}
\label{eq:HSb}
\displaystyle\kappa\frac{dT_{\rm{rad}}}{dr}+\left(\dfrac{r_i}{r}\right)^2F_{\rm{rad}}=-\dfrac{1}{r^2}\int_{r_i}^r H_s(r') r'^2 dr',
\end{equation}
showing that we can generate any functional form we desire for $dT_{\rm{rad}}/dr$ with a suitable choice of $H_s(r)$. Note that in practice (see below), the exact expressions for $H_s(r)$ and $T_{\rm{rad}}(r)$ are not needed.

We non-dimensionalize the problem by using $[l]=r_o$, $[t]=r_o^2/\nu$, $[u]=\nu/r_o$ and $[T]=|dT_o/dr-dT_{\rm{ad}}/dr| r_o$ as the unit length, time, velocity and temperature respectively, where ${dT_o}/{dr}\equiv{dT_{\rm{rad}}}/{dr}|_{r=r_o}$ is the radiative temperature gradient at the outer boundary. Then, we can write the non-dimensional equations as:
\begin{equation}
\label{eq:divU}
{\nabla}\cdot{\vel}=0,
\end{equation}
\begin{equation}
\displaystyle\frac{\partial{\vel}}{\partial{t}}+{\vel}\cdot{\nabla}{\vel}=-{\nabla}{ p}+\frac{\text{Ra}_o}{\text{Pr}}{\Theta}\boldsymbol{e_r}+{\nabla}^2{\vel},
\end{equation}
and
\begin{equation}
\label{eq:heq}
\displaystyle\frac{\partial{\Theta}}{\partial{ t}}+{\vel}\cdot{\nabla}{\Theta}+\beta({r}){u_r}=\frac{1}{\rm{Pr}}{{\nabla}^2{\Theta}}.
\end{equation}
In all that follows, all the variables and parameters are now implicitly non-dimensional. This introduces  the Prandtl number Pr and the global Rayleigh Ra$_o$ defined as
\begin{equation}
\text{Pr}=\displaystyle\frac{\nu}{\kappa}\quad \text{and}\quad \text{Ra}_o=\displaystyle\frac{\alpha g\left|\displaystyle\frac{dT_o}{dr}-\frac{dT_{\rm{ad}}}{dr}\right|r_o^4}{\kappa\nu},
\label{eq:Rao}
\end{equation}
as well as the function $\beta(r)$ which is given by 

\begin{equation}
\beta(r)=\displaystyle\frac{\displaystyle\frac{dT_{\rm{rad}}}{dr}-\displaystyle\frac{dT_{\rm ad}}{dr}}{\displaystyle\left|\frac{d{T}_o}{dr}-\displaystyle\frac{dT_{\rm ad}}{dr}\right |}.
\end{equation}
By suitably selecting $H_s(r)$, and therefore $T_{\rm{rad}}(r)$, we can  create a profile for $\beta(r) $ that results in a  convectively stable region for $r_i\leq r<r_t$ and an unstable region for $r_t\leq r\leq r_o$. Here, we choose to impose a function $\beta(r)$ of the form
\begin{equation}
\label{eq:betaf}
\beta(r)=\left\{\begin{array}{l} 
\displaystyle -S\tanh\left(\frac{r-r_t}{d_{in}}\right) \mbox { when  }  r<r_t , \\
\displaystyle -\tanh\left(\frac{r-r_t}{d_{out}}\right) \mbox { when  } r\geq r_t ,
\end{array}\right.
\end{equation}
where $d_{in}$ and $d_{out}$ constrain the width of the imposed radiative-convective boundary, while $S$ is the stiffness parameter which measures the relative stability of the radiative and the convective zones. 
Note that $d_{in}$ is chosen such that the derivative of $\beta(r)$ is continuous at $r_t=0.7$, which implies that $d_{in}=Sd_{out}$. The quantity $d_{out}^{-1}$ is the derivative of the function $\beta$ at $r = r_t$, and  therefore describes the steepness of its profile.  In this model $\beta(r)$ tends to $-1$ in the bulk of the convection zone, and to $S$ in the bulk of the radiative zone. In stars, this is of course not the case, and $\beta(r)$ can vary very significantly within both convective and radiative zones, so this model is chosen for simplicity but with the ability to explore certain questions raised in the introduction related to the effect of the stiffness and the abruptness of the transition. Note that in the Sun, $|\beta(r)|$ decreases substantially from the top to the base of the convection zone \citep[see][]{Korre}, the slope of the transition into the radiation zone is rather smooth, and $|\beta(r)|$ in the radiative zone is of the same order as $|\beta|$ in the bulk of the convection zone (which implies that $S$ would be of order unity).
\par
\begin{figure}
\centering
\includegraphics[scale=0.38]{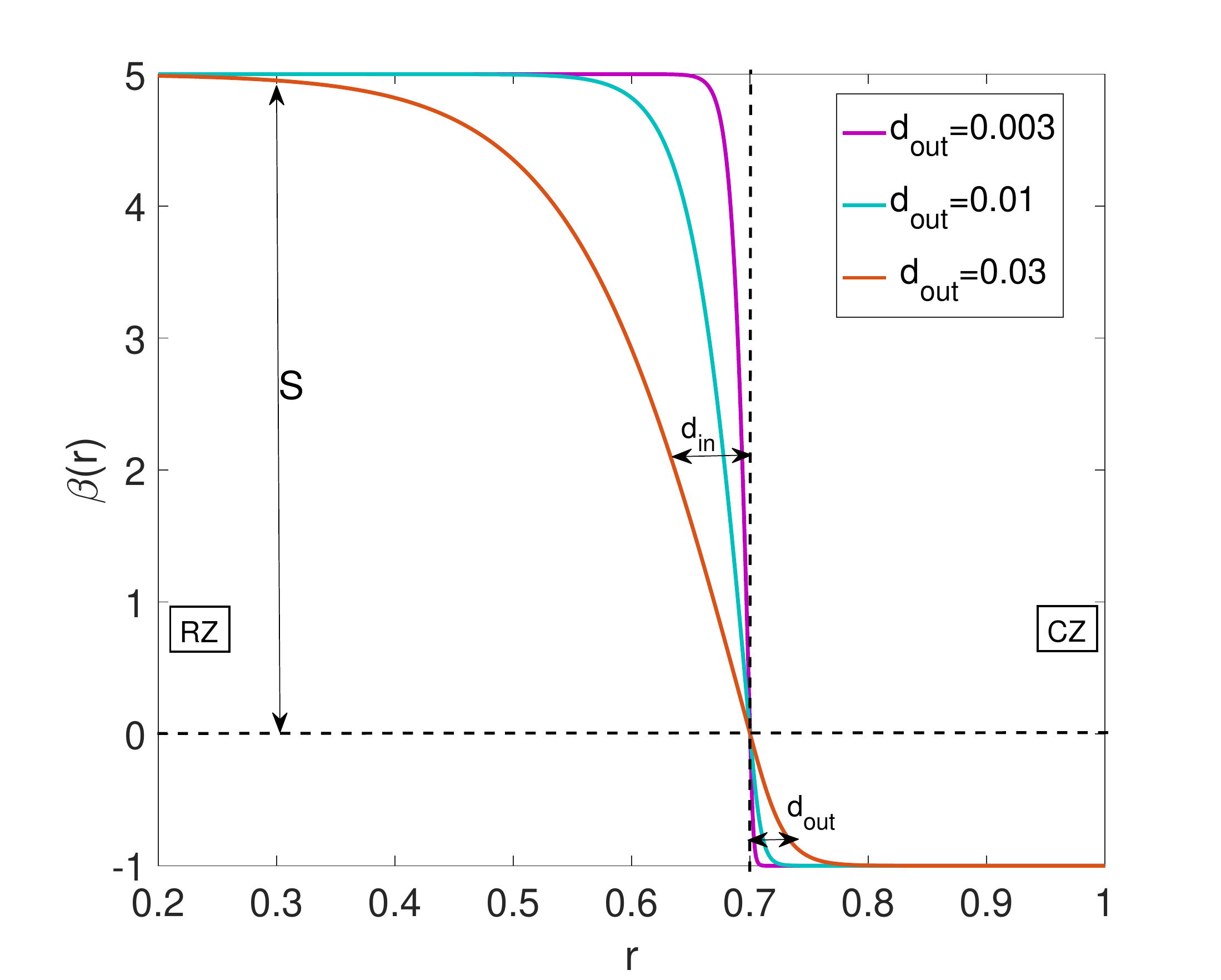}
\hspace{-13mm}
\caption{The profile of $\beta(r)$ versus the radius $r$, for $S=5$  and three different $d_{out}$ values.}
\label{fig:Fig_1}
\end{figure}

 The function  $\beta$ can also be expressed  as minus the ratio of the local Rayleigh number Ra$(r)$ to Ra$_o$, namely
\begin{equation}
\displaystyle\beta(r)=-\frac{\rm{Ra}(r)}{{\rm Ra}_o},
\end{equation}
where 
\begin{equation}
\label{eq:betaRa}
\text{Ra}(r)=-\displaystyle\frac{\alpha g\left(\displaystyle\frac{dT_{\rm{rad}}}{dr}-\frac{dT_{\rm{ad}}}{dr}\right)r_o^4}{\kappa\nu},
\end{equation}
and where the minus sign in Equation (\ref{eq:betaRa}) ensures that Ra$(r)$ is positive in convective regions.  
In Figure \ref{fig:Fig_1}, we show representative profiles of $\beta(r)$   in order to demonstrate their dependence on the two parameters $S$ and $d_{out}$.    
Higher values of $S$ result in a larger jump in $\beta(r)$ from the base of the CZ inward, while lower values of $d_{out}$ at fixed $S$ lead to a steeper and  more sudden transition.

In order to study the dynamics of our two-layered system and understand the mixing processes that occur due to the propagation of the convective motions into the stable layer, we have run 3D direct numerical simulations  (DNS) solving the Boussinesq equations in a spherical shell, exactly as outlined above, using the PARODY code \citep{Parody2,Parody1}. In all of our simulations, the Prandtl number is fixed and  equal to Pr$=0.1$. The boundary conditions for the temperature are such that we have fixed flux  at the inner boundary which translates into a no-flux boundary condition for the perturbations $\Theta$,  $\partial \Theta/\partial r|_{r_i}=0$, and fixed temperature  at the outer boundary which translates into a zero temperature boundary condition for $\Theta$, $\Theta(r_o)=0$. For the velocity, we employ stress-free boundary conditions. Each simulation is evolved from a zero initial velocity and  small-amplitude perturbations in the  temperature field until a statistically stationary and thermally-relaxed state is achieved. To determine when this is the case, we look both at the total kinetic energy per unit  volume in the domain, $E(t) = \frac{1}{2}(u_r^2+u_{\theta}^2+u_{\phi}^2)$, and at the temperature perturbation gradient at the surface.

We have run a large number of simulations, whose input parameters are summarized in Table \ref{tab:data}. In Section \ref{sec:samplesim}, we present an in-depth study of a typical simulation, focusing on identifying measures of the  dynamics of overshooting and/or penetrative convection in the vicinity of the CZ-RZ interface. In Sections \ref{sec:kinprof} and \ref{sec:thermal} we then look in turn at selected properties of our results across all available simulations.

\begin{table*}
\caption{Columns 1-7: Summary of all input parameters and resolution of our simulations. Column 8 reports on the lengthscale $\delta_G$ discussed in Section \ref{sec:samplesim}. Column 9 reports on the lengthscale $\delta_{en}$ discussed in Section \ref{sec:kinprof}.  Column 10 reports on the lengthscale $\delta_{\Theta}$ discussed in Section \ref{sec:samplesim}, and column 11 reports on $\delta_u$ discussed in Section \ref{sec:samplesim}. Columns 12 and 13 report on Pe and Pe$_{ov}$ as discussed in Section \ref{sec:samplesim}. The respective Reynolds numbers are Re$=10$Pe and Re$_{ov}=10$Pe$_{ov}$.}
\label{tab:data}
\begin{tabular}{lccccccccccccc}
\hline
Case \# & $S$ &  $d_{out}$ & $d_{in}$ & Ra$_o$ & Pr &  $N_r\times N_{\theta}\times N_{\phi}$ & $\delta_G$  & $\delta_{en}$ & $\delta_{\Theta}$ &  $\delta_{u}$  & Pe & Pe$_{ov}$\\
\hline
1 &	100	& 0.0003 & 0.03 &	$10^6$ & 0.1 & 350$\times$192$\times$192 & 0.020 &	0.017 & 0.059	& 0.048 & 7.3 &	0.5

\\
2 & 5	& 0.003 & 0.015	& $10^6$ & 0.1 & 300$\times$192$\times$192 & 0.060	 & 0.045 & 0.150	& 0.130 & 8.4 & 1.7

\\
3 & 10 & 0.003	& 0.03	& $10^6$ & 0.1 & 300$\times$192$\times$192 & 0.049	& 0.039 & 0.130	& 0.100 & 8.2 &	1.3
\\
4	& 20 &	0.003 &	0.06 & $10^6$ & 0.1 & 300$\times$192$\times$192 & 0.044 &	0.037 & 0.110 &	0.092 &8.2	& 1.2
\\
5 &	2 &	0.01 &	0.02 & $10^6$ & 0.1 & 300$\times$192$\times$192 & 0.096 &	0.069 & 0.240 &	0.220 & 8.9 &	2.9
\\
6 &	5 &	0.01 & 0.05	& $10^6$ & 0.1 & 300$\times$192$\times$192 &  0.071 &	0.057 & 0.170 &	0.150 & 8.8 &	2.1
\\
7	& 10 &	0.01 & 0.1	& $10^6$ & 0.1 & 300$\times$192$\times$192 & 0.064 &	0.054 & 0.150	& 0.130 & 8.6 &	1.8
\\
8 &	5 &	0.03 &	0.15 & $10^6$ & 0.1 & 300$\times$192$\times$192 & 0.091 &	0.078 & 0.220  & 0.190 & 9.0 &	2.7
\\
9 &	10	& 0.03 & 0.3 & $10^6$ & 0.1 & 300$\times$192$\times$192 & 0.089 &	0.077 & 0.210	& 0.190 & 9.0 & 2.6
\\
10 & 5	& 0.05 & 0.25 & $10^6$ & 0.1 & 300$\times$192$\times$192 & 0.104 &	0.091 & 0.260	& 0.230 & 9.0 &	3.1
\\
11 & 5 & 0.003 & 0.015	& $10^7$ & 0.1 & 400$\times$288$\times$320 & 0.047 & 0.034 & 0.110	& 0.094 & 21.7 & 	3.4
\\
12	 & 10 &	0.003 &	0.03 & $10^7$ & 0.1 & 400$\times$288$\times$320 & 0.038 &	0.030 & 0.090 &	0.075 & 21.7 &	2.7
\\
13	& 20 & 0.003 & 0.06	& $10^7$ & 0.1 & 400$\times$288$\times$320 & 0.035 & 0.029 & 0.085	& 0.070 & 21.5	& 2.5
\\
14 & 5 & 0.01 &	0.05 & $10^7$ & 0.1 & 400$\times$288$\times$320 & 0.057 &	0.045 & 0.140 & 0.120 & 22.3 &	4.2
\\
15 & 10	& 0.01	& 0.1 & $10^7$ & 0.1 & 400$\times$288$\times$320 &	0.052 &	0.043 & 0.120	& 0.104 & 22.0 & 3.8
\\
16	& 5	& 0.03	& 0.15	& $10^7$ & 0.1 & 400$\times$288$\times$320 & 0.077 &	0.062 & 0.170	& 0.150 & 22.6 & 5.8
\\
17 &	10 &	0.03 &	0.3	& $10^7$ & 0.1 & 400$\times$288$\times$320 &	0.075 & 0.062 & 0.170	& 0.150 & 22.5 & 5.6
\\
18	& 5	 & 0.05 &	0.25	& $10^7$ & 0.1 & 400$\times$288$\times$320 &	0.090	 & 0.073 & 0.200 &	0.180 & 22.1 &	6.6
\\
19	& 5	& 0.003	& 0.015	& $10^8$ & 0.1 & 585$\times$516$\times$640 & 0.035 &	0.026 & 0.080 & 0.069 & 50.5 &	5.9
\\
20	& 10 &	0.003 &	0.03	& $10^8$ & 0.1 & 585$\times$516$\times$640 &	0.030 &	0.024 & 0.068 &	0.058 & 49.7 & 4.9
\\
21 &	20	& 0.003 &	0.06	& $10^8$ & 0.1 & 585$\times$516$\times$640 &	0.028 &	0.024 & 0.063	& 0.054 & 49.6 & 4.6
 \\
22 &	5 &	0.01 &	0.05	& $10^8$ & 0.1 & 585$\times$516$\times$640 &	0.045 &	0.036 & 0.097 &	0.089 & 51.6 &	7.8
\\
23 & 10 & 0.01 & 0.1	& $10^8$ & 0.1 & 585$\times$516$\times$640 & 0.043 &	0.035  & 0.094 &	0.085 & 51.5 &	7.3
\\
24 &	5 &	0.03 &	0.15	& $10^8$ & 0.1 & 585$\times$516$\times$640 &	0.063 &	0.050 & 0.140 &	0.120 & 51.5 &	10.9
\\
25 &	10 &	0.03 &	0.3	& $10^8$ & 0.1 & 585$\times$516$\times$640 &	0.062 & 0.050 & 0.130	& 0.120 & 51.3 & 10.6
 \\
\hline
\end{tabular}
\end{table*}
 
\section{GENERAL CHARACTERISTICS OF A TYPICAL SIMULATION}
\label{sec:samplesim}

\par Throughout the paper, we define the time- and spherical- average of a quantity as   
\begin{equation}
\bar{q}(r)=\displaystyle\frac{1}{4\pi(t_2-t_1)}\int_{t_1}^{t_2}\int_0^{2\pi}\int_0^{\pi} q(r,\theta,\phi,t)\sin\theta d\theta d\phi dt,
\end{equation}
where $t_1$ and $t_2$ are an initial and a final time, taken once the system has reached a statistically stationary state. 
We sometimes choose to present properties of the downflows and upflows separately. Therefore, we also define the average over downflows and upflows only as 

\begin{equation}
\bar{q}_{down}(r) = \dfrac{1}{A_{down} (t_2-t_1)} \int_{t_1}^{t_2}\int_0^{2\pi}\int_0^{\pi} q(r,\theta,\phi,t) H(-u_r)  \sin \theta d \theta d\phi dt,
\end{equation}

\begin{equation}
\bar{q}_{up}(r) = \dfrac{1}{A_{up} (t_2-t_1)} \int_{t_1}^{t_2}\int_0^{2\pi}\int_0^{\pi} q(r,\theta,\phi,t) H(u_r) \sin \theta d \theta d\phi dt,
\end{equation}
where $H$ is the Heaviside function, $A_{down}$ is the area covered by the downflows, namely 
\begin{equation}
A_{down}(r)=\dfrac{1}{t_2-t_1} \displaystyle\int_{t_1}^{t_2}\int_0^{2\pi}\int_0^{\pi} H(-u_r)\sin\theta d\theta d\phi dt,
\end{equation}
and $A_{up}$ is the area of the upflows such that 
\begin{equation}
A_{up}(r)=\dfrac{1}{t_2-t_1} \displaystyle\int_{t_1}^{t_2}\int_0^{2\pi}\int_0^{\pi} H(u_r)\sin\theta d\theta d\phi dt.
\end{equation}

We begin by presenting the results of a typical run where $S=5$, $d_{out}=0.003$, and Ra$_o=10^7$ (Case 11 in Table \ref{tab:data}), which illustrates some of the most basic characteristics observed in almost all of our simulations. Table \ref{tab:data} summarizes its input parameters, resolution, and some of the quantities of interest discussed below. The profile of $\beta(r)$ corresponding to these parameters is shown as the purple line in Figure \ref{fig:Fig_1}.

\begin{figure*}
\centering

\includegraphics[scale=0.18]{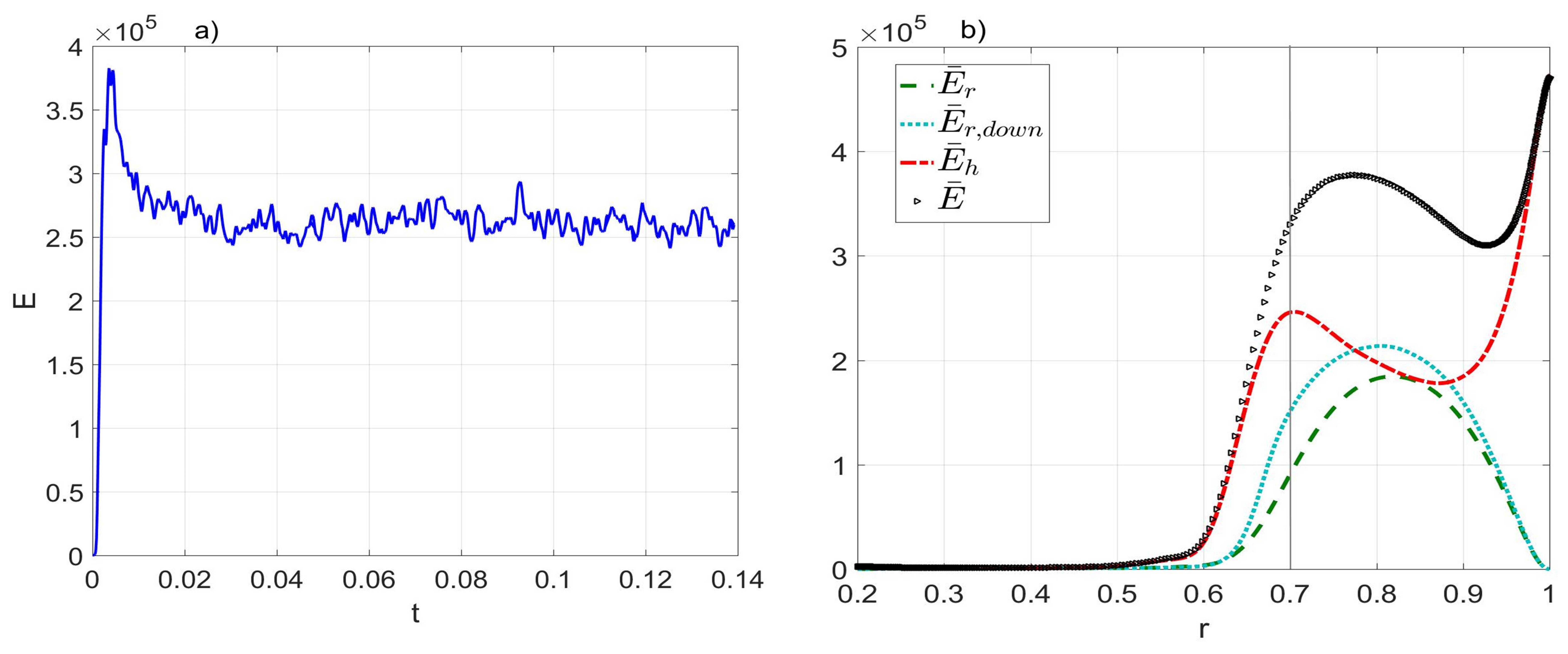}
\caption{\textit{a)} Non-dimensional kinetic energy per unit volume as a function of time for a typical simulation with $S=5$, $d_{out}=0.003$ and Ra$_o=10^7$.  \textit{b)} Time-averaged non-dimensional kinetic energy profiles as a function of radius, for the same simulation, where $\bar{E}_r$ is the total kinetic energy (black triangles),  $\bar{E}_{r}$ is the radial component of the kinetic energy (dashed green line), $\bar{E}_{r,down}$ is the vertical kinetic energy of the downflows (dotted cyan line), and $\bar{E}_h$ is the horizontal component of the kinetic energy (red line).}
\label{fig:Fig_2}
\end{figure*}

Figure \ref{fig:Fig_2}(a) shows the evolution of the total kinetic energy per unit volume $E$ as a function of time $t$ in the simulation. We observe the initial development of the convective  instability   in the interval  $t\in[0,0.01]$ as a large spike,  followed by  its  nonlinear saturation. The system reaches a statistically steady state in this global quantity very fast because the energy is dominated by the dynamics of the CZ which rapidly equilibrates.
However,   we must also make sure that the system reaches global thermal equilibrium. This   occurs on a much slower timescale, which depends on the radiative diffusion through the RZ. In our simulations, we estimate that this has occurred when $\partial \Theta/\partial r |_{r= r_o} $ is statistically stationary and close to zero. This happens around $t=0.04$ in this case.
   \par
\begin{figure*}
\centering
\includegraphics[scale=0.7]{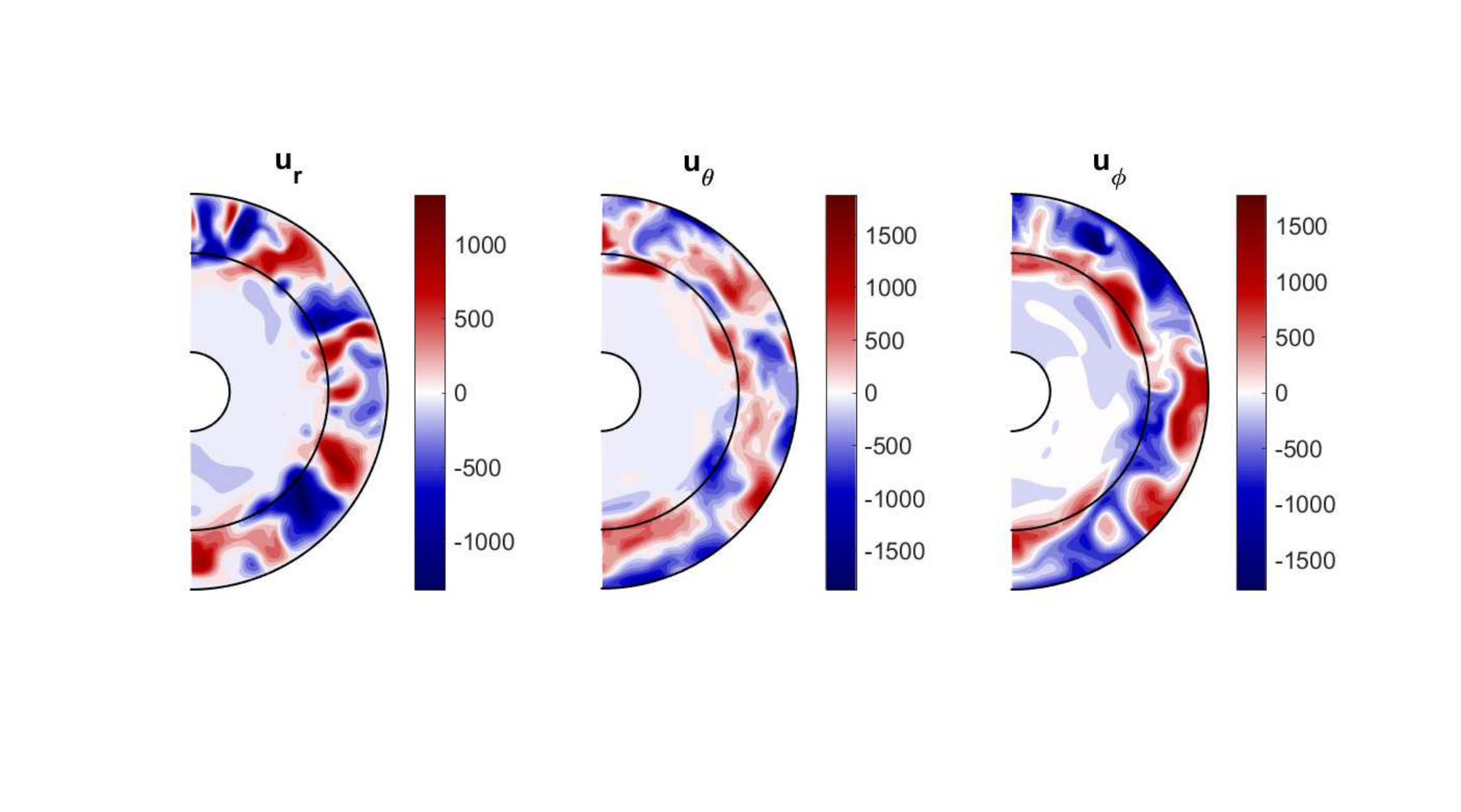}
\caption{Snapshot slice showing the velocities $u_r$, $u_{\theta}$ and $u_{\phi}$ on a selected meridional plane for a typical simulation of $S=5$, $d_{out}=0.003$ and Ra$_o=10^7$. The inner black line represents the base of the convection zone at $r_t$.}
\label{fig:Fig_3}
\end{figure*}

  In Figure \ref{fig:Fig_3}, we present   snapshots of  meridional slices of the velocity components as a function of depth and latitude, for a selected longitude, all taken at the same time $t$ during the statistically stationary state. They clearly show that the convective motions driven within the CZ are not confined to that region, but instead, travel some distance beyond the CZ-RZ interface (marked by the  inner black line at $r_t=0.7$).

\noindent As we shall demonstrate, there are many ways in which one can quantitatively study the effect of convective motions which overshoot below the base of the CZ, such as through their kinetic energy, through their effect on the mean temperature profile, as well as  through their vertical coherence.  Each of these diagnostics presents a different facet of the problem, that we will  try to reconcile through modeling in the following Sections.

We begin with  Figure \ref{fig:Fig_2}(b) which  shows  the non-dimensional kinetic energy profile $\bar{E}(r)$  given by
   \begin{eqnarray}
   \label{eq:sphKE}
   \bar{E}(r)=\displaystyle\frac{1}{2}\overline{({u_r^2+u_{\theta}^2+u_{\phi}^2)}},
   \end{eqnarray}
   as black triangles.
   On the same figure, we plot the radial component of the kinetic energy (dashed green line), the  vertical kinetic energy of the downflows (dotted cyan line), as well as the horizontal component of the kinetic energy (red line) given respectively by
 \begin{eqnarray}
   \label{eq:sphKEall}
  \bar{E}_{{r}}(r)=\displaystyle\frac{1}{2}\overline{u_r^2},\quad  \bar{E}_{{r,down}}(r)=\displaystyle{\frac{1}{2}\overline{u_{r}^2}_{,down}}, \quad \text{and}\nonumber \\
   \bar{E}_{{h}}(r)=\displaystyle\frac{1}{2}\overline{({u_{\theta}^2+u_{\phi}^2)}}.
   \end{eqnarray}
There is clearly significant kinetic energy below the CZ corresponding to overshooting. Below the CZ, the motions are no longer convectively driven and must decelerate. This causes $\bar{E}(r)$ to decrease sharply inward from the base of the convection zone. Furthermore, we see that the contributions to $\bar{E}(r)$ coming from radial and horizontal motions behave very differently from one another. The vertical kinetic energy $\bar E_r$ peaks in the middle of the CZ, and then decreases inward, a result we  attribute to a deceleration of the downflows as they approach the CZ-RZ interface at $r_t=0.7$ from above. This can indeed be verified in the profile of $\bar{E}_{r,down}$ which has the same properties, although we also see that it is a little larger, indicating  that downflows must be on average stronger (but narrower) than the upflows (this can be verified by a direct inspection of $A_{up}$ and $A_{down}$, not shown). Meanwhile, the horizontal kinetic energy increases substantially near the bottom of the convection zone.  Thus, there is an exchange of kinetic energy between the vertical  and the horizontal flows, which we interpret as the result of a deflection of the vertical plumes towards the horizontal.  While this may seem somewhat expected, it is interesting to see that this occurs in the bulk of the CZ and not only near or below the CZ-RZ interface,  implying that the presence of this interface is felt in a highly non-local way throughout the entire convection zone. This result is not an artifact of the Boussinesq approximation, since it is also seen in anelastic and fully compressible 2D simulations  \citep[e.g.][]{Rogers2005,Pratt17} and in 3D fully compressible simulations \citep[e.g.][]{Singh95,Brummell}.

 From the CZ-RZ interface downward, we observe  a rapid decrease in $\bar{E}(r)$, which is expected from the stabilizing effect of the stratification.
 Note that  that since the energy in the vertical motions has  already decreased significantly even before reaching the CZ-RZ interface, most of the remaining energy below the base of the CZ comes from horizontal motions only. This leads to the conclusion that horizontal motions are dominant in the average sense below the CZ and therefore have to be considered in the study of convective overshooting dynamics,  as in the models of e.g. \citet{vanBallegooijen} and \citet{Rempel2004}.

\par
\begin{figure}
\centering
\includegraphics[scale=0.32]{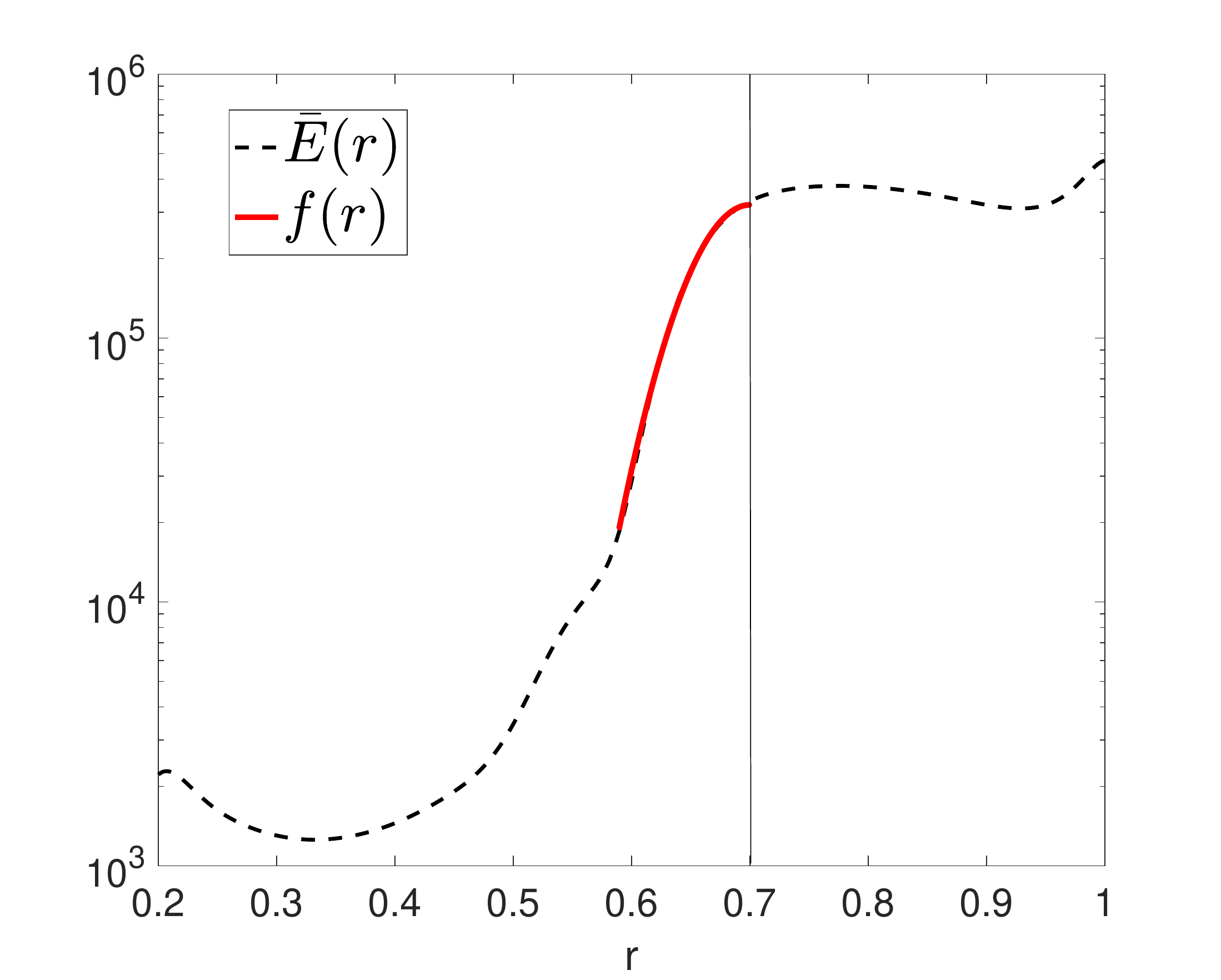}
\caption{Kinetic energy profile $\bar{E}(r)$  for $S=5$, $d_{out}=0.003$ and Ra$_o=10^7$ against the radius $r$. The red solid line is the fitted curve of the kinetic energy profile on this interval.}
\label{fig:Fig_4}
\end{figure}

 In Figure \ref{fig:Fig_4}, we plot the total kinetic energy $\bar{E}(r)$ on a log scale to clarify its features below $r_t$. We see that $\bar{E}(r)$ drops significantly faster than exponentially with depth below the CZ in contrast with the model proposed by  \citet{Freytag96} \citep[also see][]{Herwig2000}. In fact, we find that a Gaussian function of the kind
\begin{equation}
f(r)=A \exp\left(-\frac{(r-r_t)^2}{2\delta_G^2}\right)
\label{eq:Gaussian}
\end{equation}
with $\delta_G = 0.047$ would be a much better fit to the profile of $\bar{E}(r)$, at least down to $r = 0.58$, as shown by the red solid line in Fig. \ref{fig:Fig_4}. Below that point, the decay of the kinetic energy is closer to being exponential in the interval $[0.47,0.58]$. Even deeper down, $\bar E(r)$ flattens out, presumably as a result of the presence of the inner boundary. The Gaussian function $f(r)$ can be used to characterize the spherically-averaged kinetic energy profile of overshooting motions below the CZ, and is parametrized by its amplitude $A$, and by its width $\delta_G$.  Therefore $\delta_G$ can be used to characterize the region of influence of convective motions in the stable RZ, at least energetically speaking and in an average sense.

\begin{figure}
\centering
\includegraphics[scale=0.35]{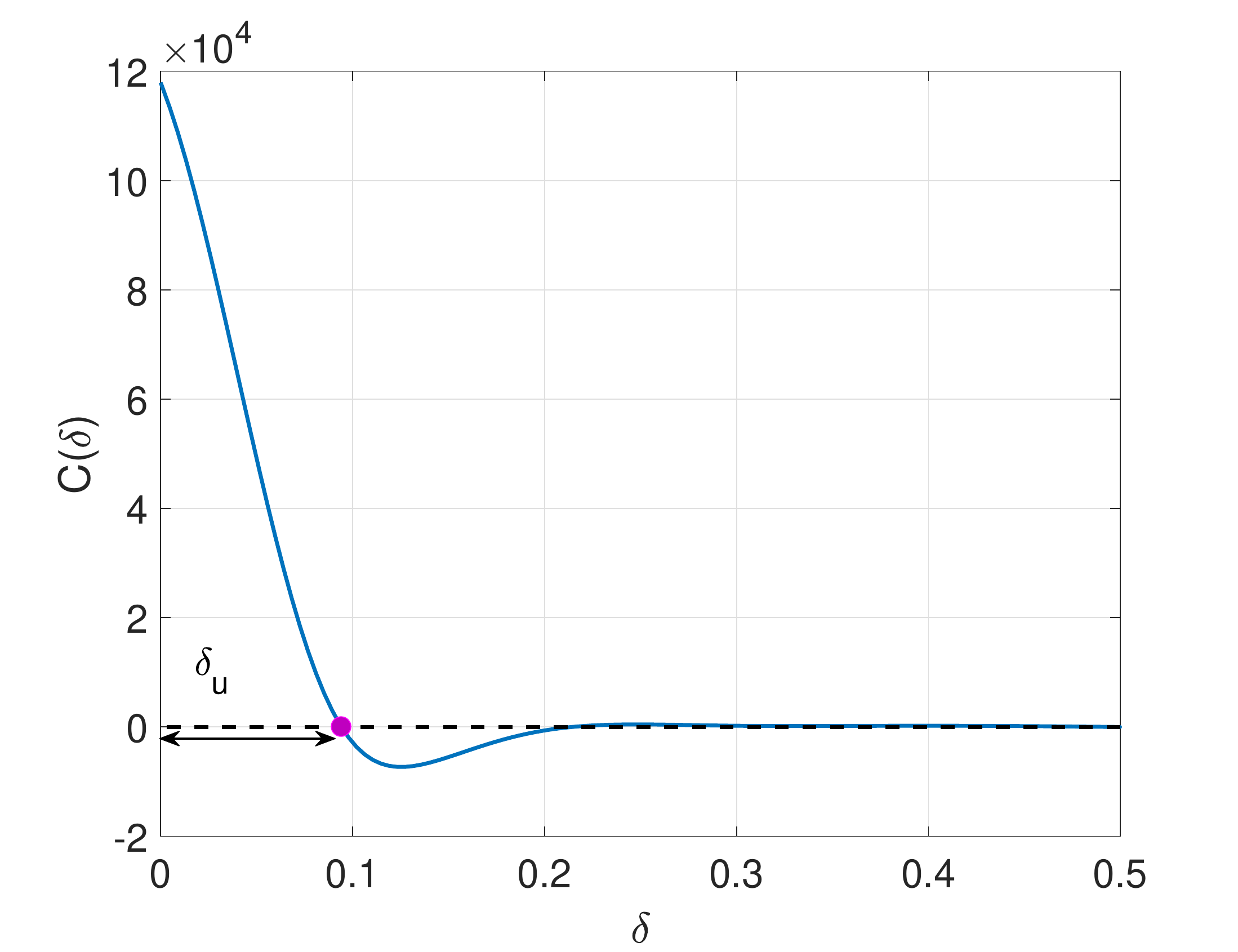}
\caption{ Profile of $C(\delta)$ against $\delta$ for $S=5$, $d_{out}=0.003$, and for  Ra$_o=10^7$.}
\label{fig:Fig_5}
\end{figure}

An alternative measure is the distance that the strongest of the downflow motions  travel into the stable region, therefore, we introduce  the radial correlation function of the vertical velocity field in the downflows
\begin{equation}
C(\delta)=\dfrac{1}{4\pi(t_2-t_1)}\int_{t_1}^{t_2}\int_0^{2\pi} \int_0^{\pi}  u_r(r_t,\theta,\phi) H(- u_r(r_t,\theta,\phi) ) u_r(r_t-\delta,\theta,\phi) \sin \theta d \theta d\phi dt.
\end{equation}
This definition clearly favors the strongest downflows. 
Figure \ref{fig:Fig_5} shows $C(\delta)$ for our reference simulation. As expected, $C$ decreases with depth $\delta$ below the base of the CZ. Interestingly, we see that instead of merely approaching zero (which would indicate a gradual loss of correlation), $C(\delta)$ actually changes sign (here at $\delta =0.094$).  This implies that (1) the strongest downflows stop, on average, at a well-defined depth below the base of the CZ and that (2) there must be an upflow below each of these downflows. This can only occur if the downflow spreads laterally upon entering the RZ, and the lateral divergence of the fluid acts as a pump for the deeper upflow. This was in fact seen in all of our simulations. We therefore define a second measure of overshooting, the correlation depth $\delta_{u}$ as the first zero of $C(\delta)$. This depth measures the average stopping distance of the strongest downflows. By comparison with Figure \ref{fig:Fig_4}, we see that $r_t - \delta_u$  corresponds to the  radius where the kinetic energy switches from the Gaussian to  the exponential profile below the CZ.  This might be expected since a radical change in the dynamics of the fluid is taking place at $r_t - \delta_u$. 

\par
\begin{figure}
\centering
\includegraphics[scale=0.35]{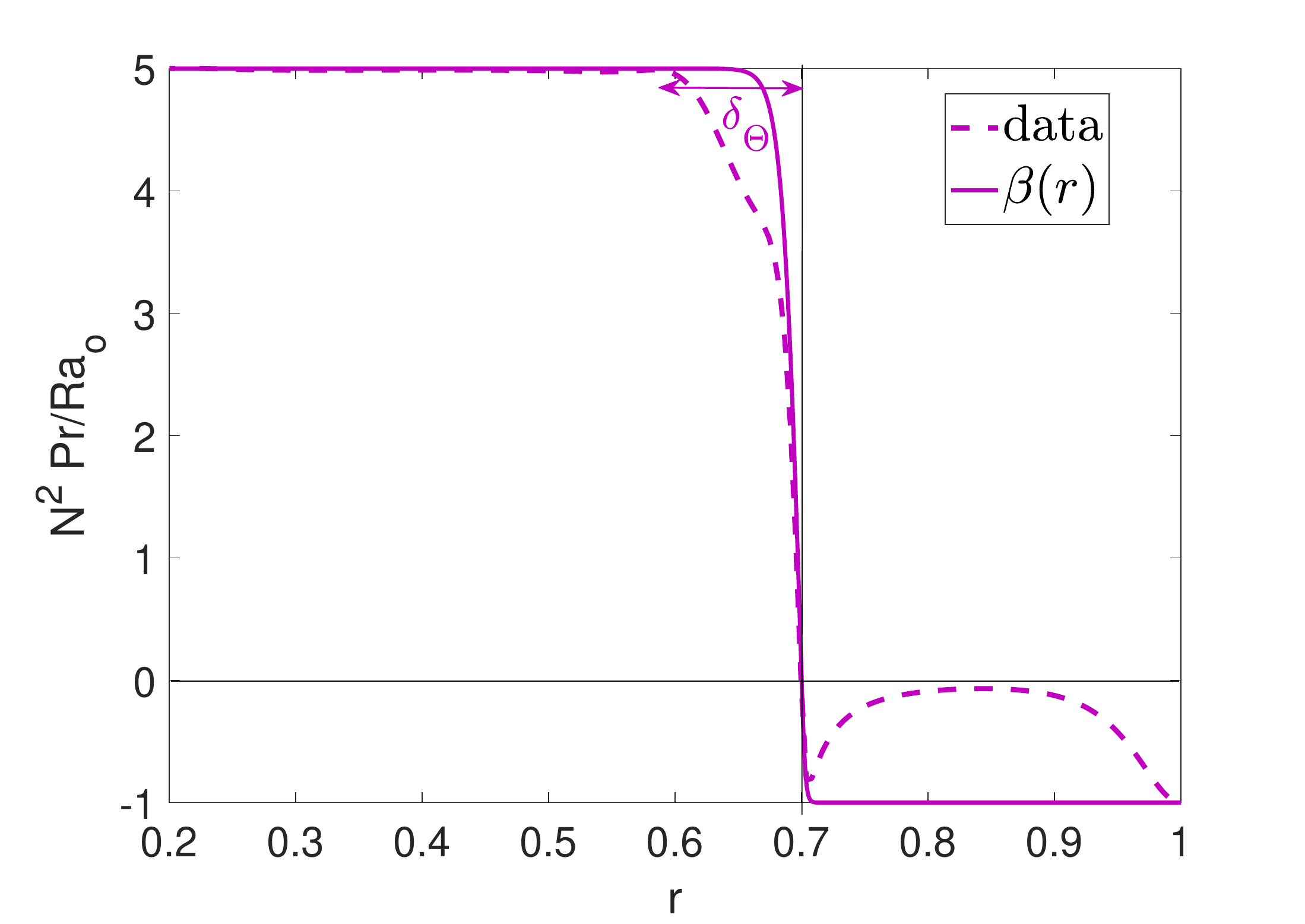}
\caption{Non-dimensional square of the buoyancy frequency $\bar{N}^2$Pr/Ra$_o$ (dashed line) compared with the background $N_{\rm{rad}}^2$Pr/Ra$_o$ (solid line) for $S=5$, $d_{out}=0.003$, and for  Ra$_o=10^7$.}
\label{fig:Fig_6}
\end{figure}

 By focusing on the fluid motions  until this point, we were only able to address the questions pertaining to overshooting rather than penetration. In order to see whether penetration occurs,  we must see if substantial thermal (entropy) mixing is occurring.  We therefore examine the non-dimensional spherically-averaged buoyancy frequency $\bar N$ whose square is given by  
\begin{equation}
\bar{N}^2(r) = \alpha g \left(\frac{d\bar{T}}{dr} - \frac{dT_{\rm{ad}}}{dr}\right) \frac{r_o^4}{\nu^2}=\left(\beta(r)+\frac{d\bar{\Theta}}{d r}\right)\frac{\text{Ra}_o}{\text{Pr}}.
\end{equation}
Figure \ref{fig:Fig_6} shows the profile of $\bar{N}^2\rm{Pr/Ra}_o$ measured in our simulations along with the original imposed background  profile $N^2_{\rm{rad}}(r){\rm{Pr/Ra}}_o=\beta(r)$ as a solid line for reference. As expected, the convective motions in the bulk of the  CZ (away from both the top boundary and  the CZ-RZ interface) mix potential temperature and drive the mean radial temperature gradient towards an adiabatic state where $\bar{N}^2\approx 0$. Below the CZ, we notice that the fluid motions do  affect the thermal stratification, but not strongly enough to effectively extend the region where $\bar{N}^2\approx 0$. This  indicates that there is no penetration  (in the strict definition of the term), but also shows that the resultant profile of $\bar{N}^2$ below $r_t$ is much smoother than the originally imposed one. This partially mixed region, which defines an intermediate state that is neither pure penetrative convection nor  pure overshooting, was found in nearly all of our simulations and this is investigated in detail in Section  \ref{sec:thermal}. This result is not entirely surprising. Indeed, the possibility of such an intermediate state was already discussed by \citet{Zahn91} and \citet{Schmitt84} (albeit briefly), and 3D fully compressible simulations to date have reported similar findings \citep[e.g.][]{Brummell,Kapyla}.

\begin{figure}
\centering
\includegraphics[scale=0.35]{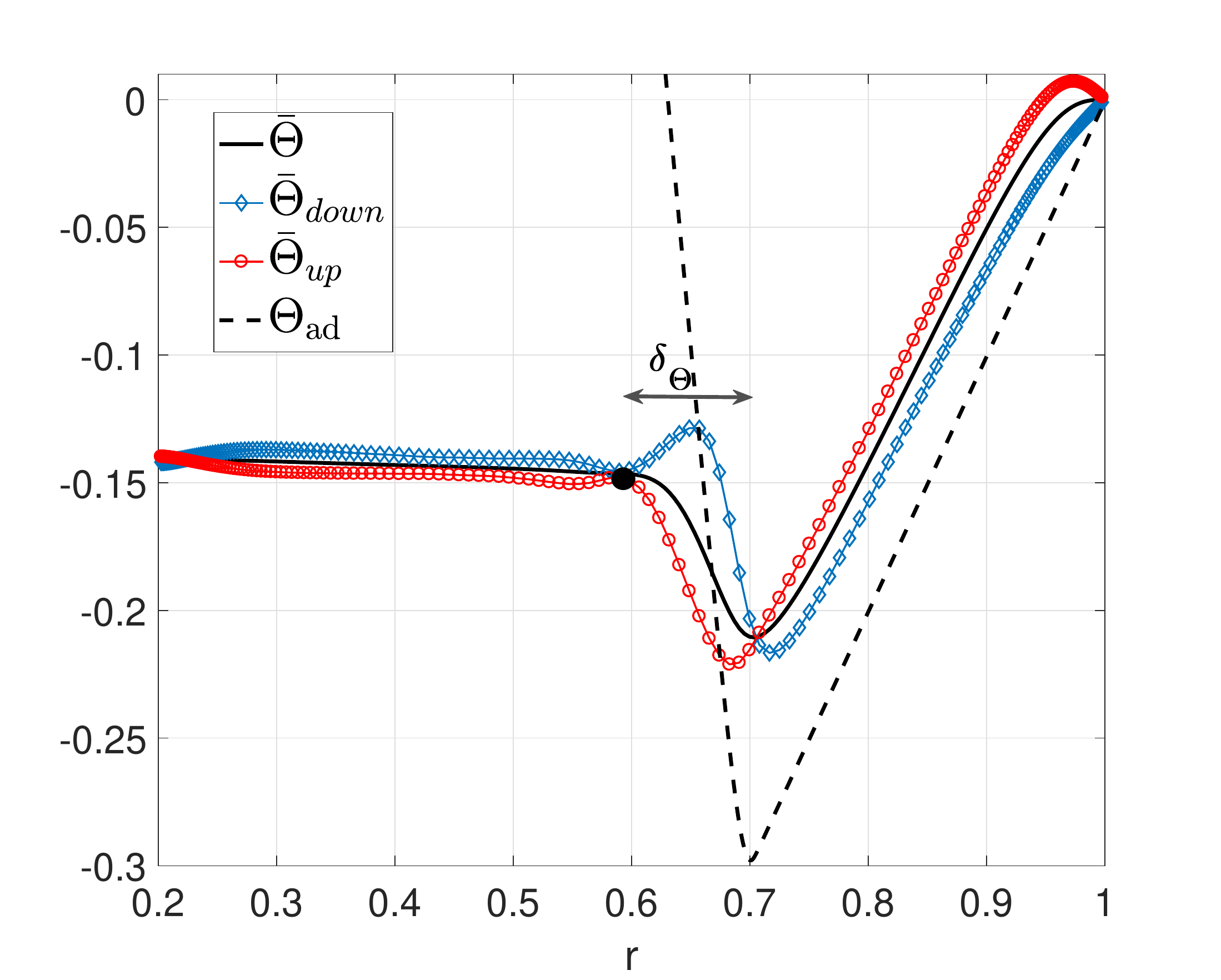}
\caption{Temperature perturbations for $S=5$, $d_{out}=0.003$, and for  Ra$_o=10^7$ plotted along with the adiabatic temperature $\Theta_{\text{ad}}$.}
\label{fig:Fig_7}
\end{figure}

To better understand what might be the cause of this partial mixing, we now look at the details of the thermal transport. Figure \ref{fig:Fig_7} shows the time- and spherically-averaged temperature perturbations $\bar{\Theta}$ along with the mean temperature perturbation in the upflows ($\bar{\Theta}_{up}$) and downflows ($\bar{\Theta}_{down}$). We also show the temperature that a downflow traveling adiabatically from  the surface (where $\Theta = 0$) would have as a function of $r$, namely $\Theta_{\text{ad}}(r)=-\int_{r_o}^r\beta(r')dr'$. 
We observe that the mean temperature gradient follows the adiabatic one quite closely in the CZ, but that $\bar \Theta$ is systematically larger than $\Theta_{\rm ad}$ due to the existence of the outer thermal boundary layer.   
Moreover, we see that $\bar{\Theta}_{down}$ is lower than $\bar{\Theta}$ in the CZ, which is expected since cooler fluid parcels are accelerated downward. Downflowing fluid parcels crossing the base of the CZ into the RZ begin to heat up through adiabatic compression, and become significantly warmer than the mean. This provides them with an upward acceleration that gradually slows them down. Upflows follow a reverse pattern, where they are warmer than $\bar{\Theta}$ in the CZ, and cooler than $\bar{\Theta}$ in the RZ. Interestingly, we find that $\bar{\Theta}_{down}$ increases by a little just above the base of the CZ, a result that could either be due to nonlinear mixing with the warmer upflows, or, to a diffusive heat flux coming from the much warmer perturbations below the base of the CZ.  

We note that there is a point lower in the RZ (here, around $r = 0.6$), at which $\bar{\Theta}$ , $\bar{\Theta}_{down}$ and $\bar{\Theta}_{up}$ approximately coincide. We therefore define a new lengthscale $\delta_{\Theta}$ which corresponds to the distance of this point from the CZ-RZ interface. Upflows and downflows are neutrally buoyant at $r = r_t - \delta_{\Theta}$. Below that level, we see that the correlation between the temperature and the direction of the flow becomes much weaker. This then implies that motion must no longer be of convective type and therefore this lengthscale is another measure of where the dynamics change character. We find  that $\delta_\Theta \simeq \delta_u$, and as mentioned before, $r_t - \delta_u$ also appears to coincide with the radius  where the kinetic energy profile $\bar{E}(r)$ transitions from a Gaussian to an exponential (see Figure \ref{fig:Fig_4}). 
Finally, we also overlay the lengthscale $\delta_{\Theta}$ on Figure \ref{fig:Fig_6} for comparison. Not surprisingly perhaps,  we observe that  $\delta_{\Theta}$ coincides with the depth of the region in the RZ where $\bar N^2$ deviates most strongly from the radiative equilibrium profile. Therefore, $\delta_{\Theta}$  provides a lengthscale that is associated with the depth of (partial) thermal mixing in the stable region.

 Another way of quantifying the  transition from the fully-mixed CZ, to the partially mixed overshoot layer, to the unmixed interior (below $\delta_{\Theta}$), is to look at the P\'{e}clet number, which is commonly defined as the ratio of advective to diffusive thermal timescales. 
In the bulk of the convection zone, the P\'{e}clet number can be estimated as 
\begin{equation}
\label{Pe}
{\rm{Pe}} =\dfrac{u_{cz} (r_o-r_t)}{\kappa} = 0.3 u_{rms}  {\rm{Pr}},
\end{equation}
where the first expression contains only dimensional quantities, and the second only contains non-dimensional ones, and where $u_{cz}$ is the typical r.m.s. velocity of convective eddies, which non-dimensionally is equal to $u_{rms} = (2E_{CZ})^{1/2}$.  
In the overshoot layer on the other hand, an appropriate lengthscale   of eddies might be $\delta_G$, and their r.m.s. velocity drops from $u_{rms}$ to $0$ over that lengthscale, so we define the P\'{e}clet number as
\begin{equation}
\label{Peov}
{\rm{Pe}}_{ov} = u_{rms} \delta_G {\rm{Pr}}. 
\end{equation} 
Note that the definition of any P\'{e}clet number is somewhat arbitrary, since there is ambiguity in the choices of the characteristic length and velocity scales.  Using our particular choices above consistently, however, we can at least compare simulations. We find that Pe $\sim 22$, and Pe$_{ov} \sim 3$, for our canonical Case 11  of $S=5$, $d_{out}=0.003$ and Ra$_o=10^7$. This finding is consistent with the expectation that the CZ is well mixed (with a large Pe), while the overshoot layer is only partially mixed (with Pe$_{ov}$ of order unity). Note that associated Reynolds numbers can be calculated as Re=Pe/Pr=10Pe (and similarly for Re$_{ov}$).

To summarize our results so far, our inspection of the dynamics observed in this simulation has suggested the definition of three distinct lengthscales that each provides a different measure of the impact of convective motions on the underlying radiative zone. The first is the width $\delta_G$ of the Gaussian function fitted to the total kinetic energy profile below the base of the CZ. This parameterizes the profile of  the decay of the turbulent kinetic energy 
with distance away from the CZ-RZ interface. The second is $\delta_u$, given by the first zero of the radial correlation function of the downflows, $C(\delta)$. This can be interpreted as the lengthscale down to which the strongest downflows travel before stopping. The third is the distance $\delta_{\Theta}$ from the base of the CZ  down to the point of neutral buoyancy where  $\bar{\Theta}=\bar{\Theta}_{down}=\bar{\Theta}_{up}$ which is both a good estimate of the stopping of motions and of  the vertical extent of  the partially thermally mixed region in the stable RZ. We have found that $\delta_G < \delta_u \simeq \delta_\Theta$ for this simulation, a result which actually holds for all of our simulations (see Table \ref{tab:data}). This suggests that while $\delta_G$ may provide an average view of the kinetic energy available for mixing below the base of the convection zone, much of that mixing is actually controlled by the strongest downflows, which overshoot much more deeply. These results are qualitatively consistent with the findings of \citet{Brummell} and \citet{Pratt17} in fully compressible simulations, suggesting that the use of the Boussinesq approximation does not dramatically alter the dynamics of overshooting convection (at least near the base of a convective region deep within a star). In the following sections, we now look more broadly at how $\delta_G$, $\delta_u$ and $\delta_\Theta$ vary with input parameters.

\section{Modeling the kinetic energy profile below the base of the CZ}
\label{sec:kinprof}

\begin{figure}
\centering
\includegraphics[scale=0.22]{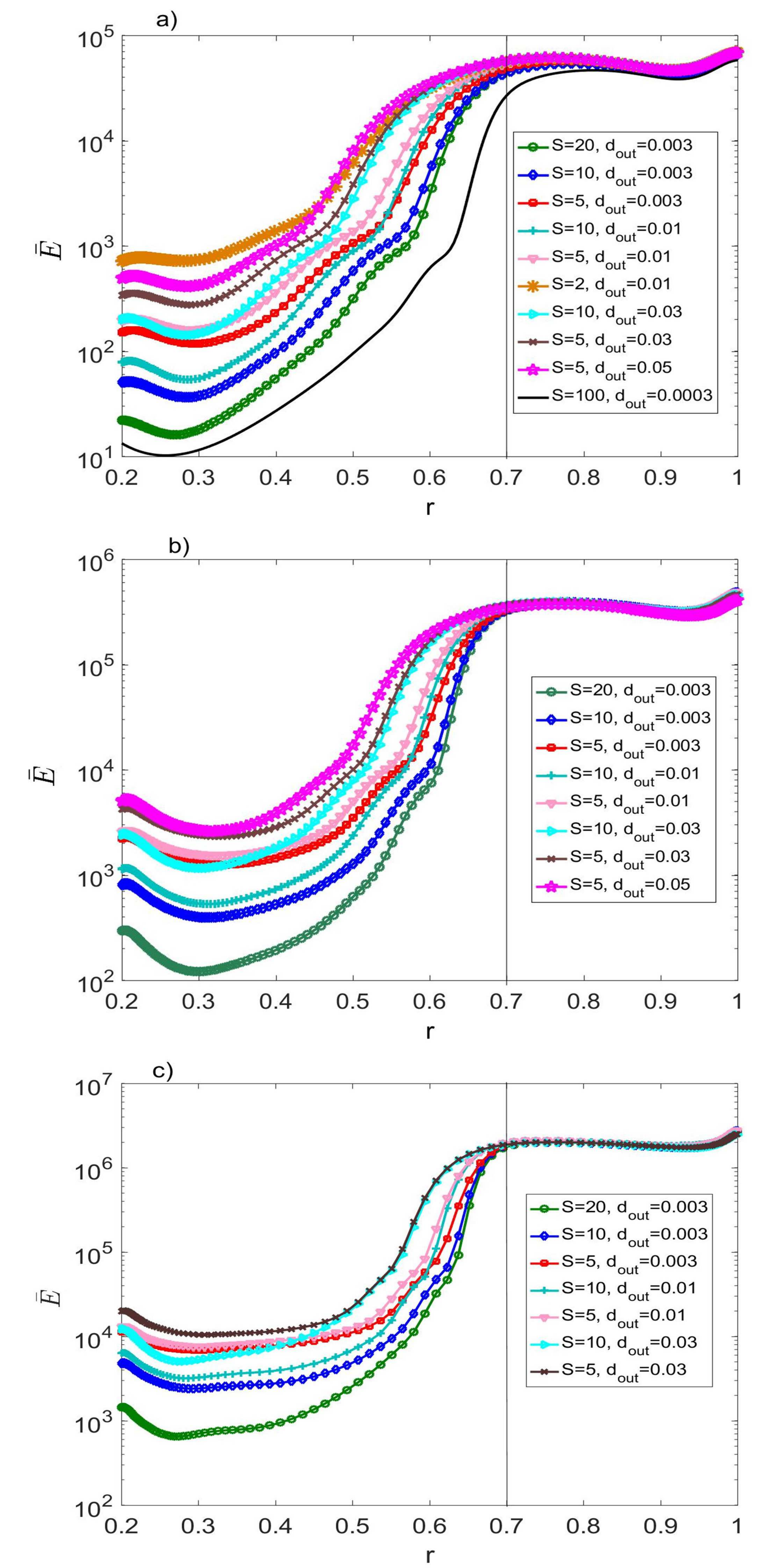}
\caption{Kinetic energy profiles on a log scale for all the different $S$, $d_{out}$ and for (a) Ra$_o=10^6$, b) Ra$_o=10^7$, and (c) Ra$_o=10^8$.}
\label{fig:Fig_8}
\end{figure}

In Section \ref{sec:samplesim}, we argued that the kinetic energy profile just below the base of the CZ resembles the Gaussian function $f(r)$ given in (\ref{eq:Gaussian}). Figure \ref{fig:Fig_8} shows this is the case in all of our simulations, which span a fairly wide range of values of the stiffness $S$, transition width $d_{out}$, and Rayleigh number ${\rm Ra}_o$. 
Comparing Figures \ref{fig:Fig_8}a, \ref{fig:Fig_8}b and \ref{fig:Fig_8}c, we clearly see that increasing the input Ra$_o$ increases the overall kinetic energy in the system (and accordingly, the amplitude of the Gaussian), which is expected since Ra$_o$ controls the strength of the convection.  Interestingly, varying $S$ and $d_{out}$ (at fixed Ra$_o$)  has very little effect on the kinetic energy within the CZ. This result is consistent with the notion that the turbulent intensity within the CZ only depends on its bulk Rayleigh number  \citep{Korre}
\begin{equation}
{\rm Ra}_b=\frac{ {\int_{r_t}^{r_o}{\rm Ra}(r)r^2 dr}}{\int_{r_t}^{r_o}r^2dr} , 
\label{eq:Rabdef}
\end{equation}
which is roughly equal to Ra$_o$ here since $\beta(r) \simeq -1$ for $r > r_t$ (see Equation (\ref{eq:betaf})).

\begin{figure}
 \centering
 \includegraphics[scale=0.3]{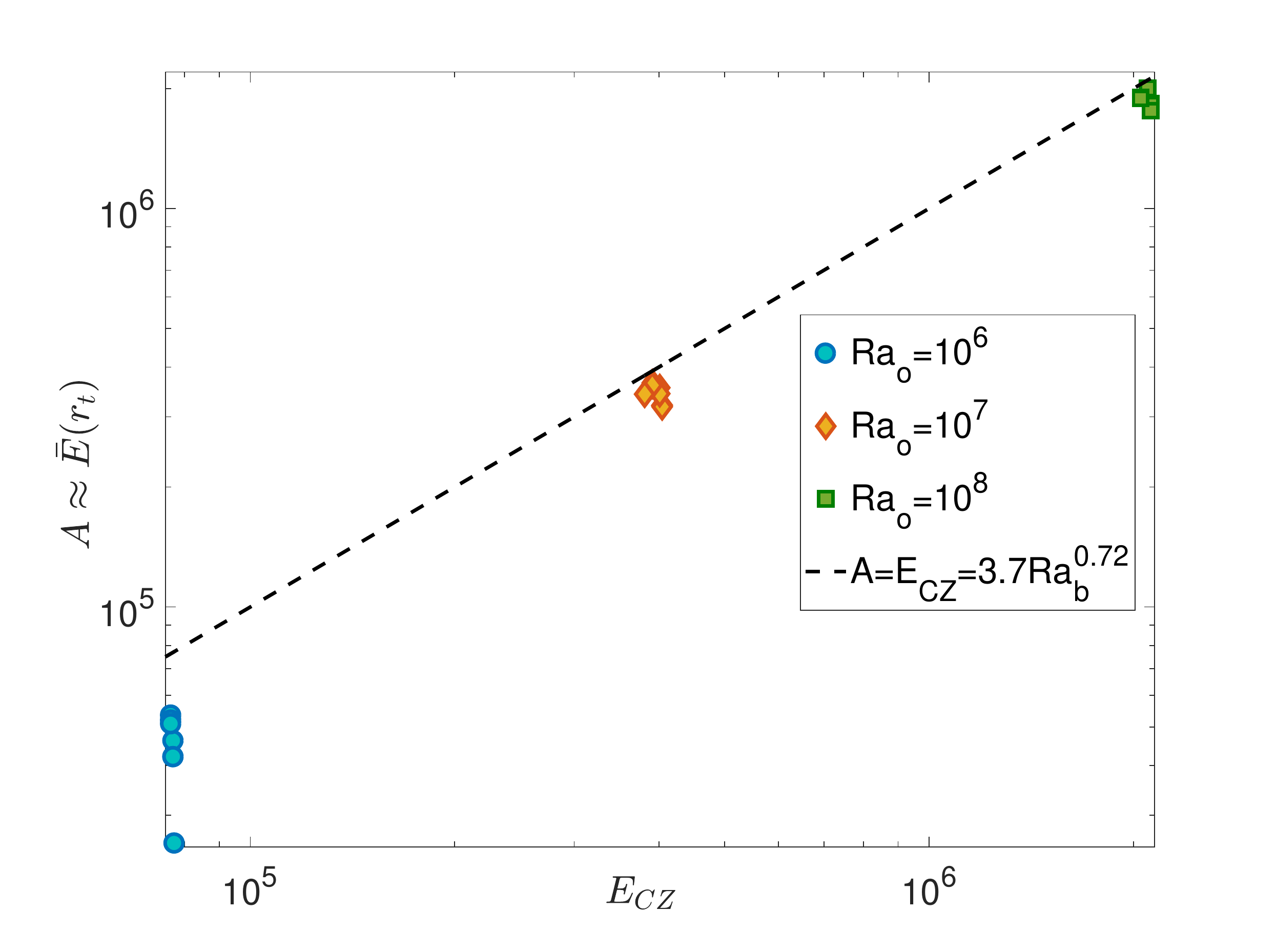}
 \caption{Plot of the extracted value of the amplitude of the Gaussian $A$, against our model for the mean kinetic energy in the CZ (see Equation (\ref{eq:ECZmod})).}
 \label{fig:Fig_9}
 \end{figure}

 \citet{Korre}  showed further that in spherical Rayleigh-B\'enard convection bounded by impermeable walls, the mean kinetic energy of the convection zone $E_{\rm CZ}$ scales as 
\begin{equation}
E_{\rm CZ} = 3.7 {\rm Ra}_b^{0.72},
\label{eq:ECZmod}
\end{equation}
when its base is at $r_t = 0.7 r_o$ and ${\rm Pr} = 0.1$, which is also the case for the CZ in this paper. To verify whether this scaling also applies in a penetrative setup and therefore could be used in a predictive model, we compare the total kinetic energy at $r_t$ to the predicted value of $E_{\rm CZ}$ in Figure \ref{fig:Fig_9}. The quantity $\bar E(r_t)$ is extracted from the simulations by fitting $f(r)$ to the data, and assuming $\bar E(r_t) \simeq A$. We see that the predicted scaling works remarkably well for the more turbulent cases (${\rm Ra}_o = 10^7$ and $10^8$), and can therefore be used to obtain a good order-of-magnitude estimate of  the amplitude of the turbulence present both within the CZ, as well as below the CZ-RZ interface through (\ref{eq:Gaussian}) provided a model for $\delta_G$ is also available.

   To construct such a model, we use a simple  energetic argument. Assuming that a parcel  travels a distance $\delta_{en}$ from the base of the CZ adiabatically down to the point where its potential energy is equal to its initial kinetic energy, we can write
\begin{equation}
\label{eq:deltapr}
E_{\rm CZ}=\delta_{en}\frac{\text{Ra}_o}{\text{Pr}}\int_{0.7-\delta_{en}}^{0.7}\beta(r) dr ,
\end{equation}
for the profile of $\beta(r)$ given in Equation (\ref{eq:betaf}). Note that this assumes that the background temperature profile has not been modified too much by the overshooting motions; we could in principle obtain a more accurate estimate for $\delta_G$ by using the actual stratification profile $\bar N^2 {\rm Pr}/{\rm Ra}_o$ computed from the simulations instead of the function $\beta(r)$ in the integrand. In practice, however, we verified that this does not make a substantial difference to the computed value of $\delta_{en}$ in any of our simulations, where thermal mixing is always weak. 
Using $\beta(r)$ in the integrand on the other hand has definite advantages: the integral can be evaluated analytically so Equation (\ref{eq:deltapr}) becomes:
\begin{align}
\label{eq:deltaen}
E_{\rm CZ}= \delta_{en}\frac{\text{Ra}_o}{\text{Pr}}Sd_{in}\ln\left[\cosh\left(\frac{\delta_{en}}{d_{in}}\right)\right].
\end{align}
Equation (\ref{eq:deltaen}) can easily be solved numerically for $\delta_{en}$, for any input $S$, Ra$_o$, and $d_{in}$. 
\par
\begin{figure}
\centering
\includegraphics[scale=0.35]{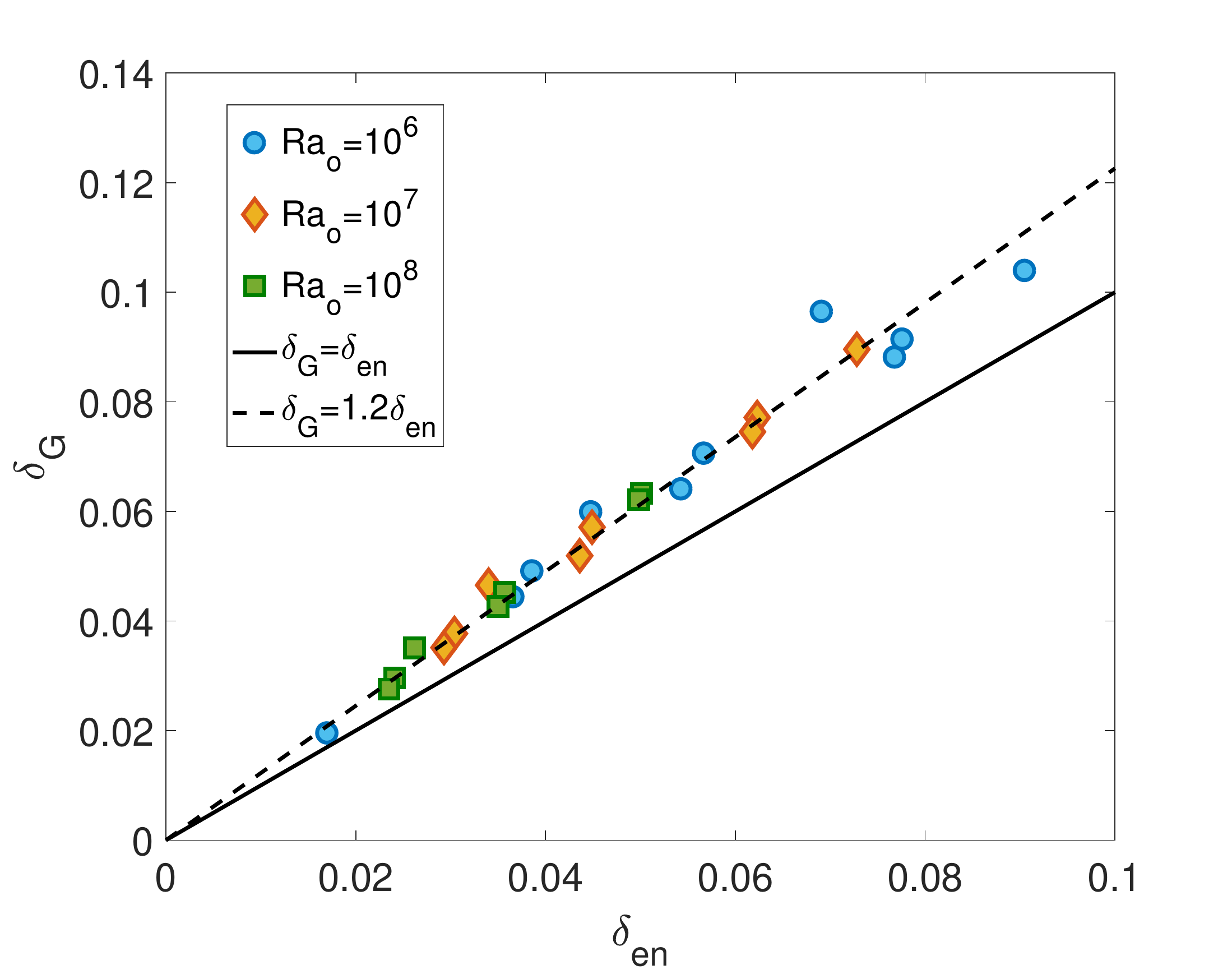}
\caption{Plot of $\delta_{en}$ versus $\delta_G$ for all the cases where Ra$_o$ has been used as  reference. \label{fig:Fig_10}}
\end{figure}

 In Figure  \ref{fig:Fig_10}, we plot $\delta_G$ against the energy-based theoretical prediction $\delta_{en}$ for all available simulations. The quantity $\delta_G$ was measured from the DNS simulations by fitting the Gaussian profile (\ref{eq:Gaussian}) to the total kinetic energy profile $\bar E(r)$ from $r_t$ down to $r_t - \delta_\Theta$ (see Section \ref{sec:samplesim}), and all the results are reported in Table \ref{tab:data}. We observe that all the points lie close to the straight line $\delta_G = 1.2 \delta_{en}$ (dashed black line). This result is rather remarkable given that our input parameters span a fairly large region of parameter space, with a resulting  $\delta_G$ ranging from $~0.01$ to $~0.12$.  
The result suggests  that the physics of the energetic argument put forward is mostly correct. Note that the downflows originating from the convection zone obviously do not all have the same kinetic energy, so $E_{\rm CZ}$ is merely an estimate of their mean, and $\delta_{en}$ is correspondingly merely an estimate of how far a typical eddy could overshoot. As a result, the prefactor relating $\delta_G$ to $\delta_{en}$ could have been any factor of order unity, but just happens to be $1.2$ in this particular set of simulations. We expect this prefactor to vary somewhat if the Prandtl number varies dramatically, or if compressibility is taken into account. 

\par
\begin{figure*}
\centering
\includegraphics[scale=0.1]{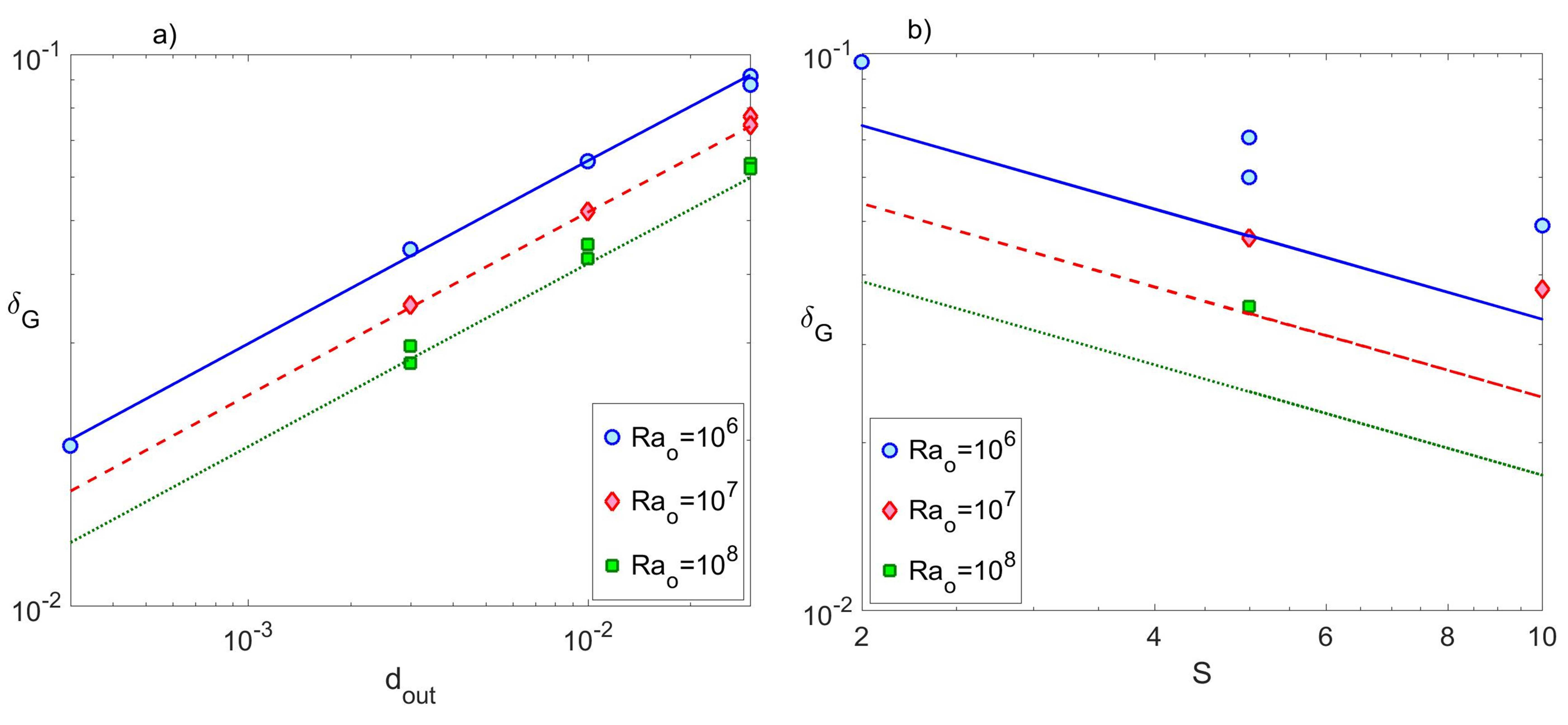}
\caption{Plot of the measured $\delta_G$ against (a) the transition width $d_{out}$, and (b)  the stiffness parameter $S$. In (a), only those simulations for which $\delta_G < \delta_{in}$ are shown. Also shown is the predicted scaling law for $\delta_{en}$ given in Equation (\ref{eq:delta1}). In (b), only those simulations for which $\delta_G > \delta_{in}$ are shown. Also shown is the predicted scaling law for $\delta_{en}$ given in Equation (\ref{eq:delta2}). }  \label{fig:Fig_11}
\end{figure*}

 While Equation (\ref{eq:deltaen}) does not have any analytical solutions in general, it has two limits of interest. When  $\delta_{en}\ll d_{in}$, Equation (\ref{eq:deltaen}) becomes
\begin{equation}
E_{\rm CZ}\simeq\delta_{en}\frac{\text{Ra}_o}{\text{Pr}}Sd_{in}\left(\dfrac{1}{2}\left(\dfrac{\delta_{en}}{d_{in}}\right)^2\right),
\end{equation} 
leading to
\begin{equation}
\label{eq:delta1}
\delta_{en}\simeq \left(\dfrac{2E_{\rm CZ}d_{out}\rm{Pr}}{{\rm Ra}_o}\right)^{1/3}\Rightarrow \delta_G\approx 1.2 \left(\dfrac{2E_{\rm CZ}d_{out}\rm{Pr}}{{\rm Ra}_o}\right)^{1/3}
=1.2 \left(\dfrac{2E_{\rm CZ}d_{in}\rm{Pr}}{{S}{\rm Ra}_o}\right)^{1/3}.
\end{equation} 
Physically speaking, this limit corresponds to the case where the downflows only sample the transition region below the CZ where $\beta(r)$ varies linearly with distance to $r_t$. As such, it is not surprising to find that $\delta_{G}$ in this case does not directly know about $S$, but only knows about the slope of $\beta(r)$.  
In Figure \ref{fig:Fig_11}(a), we plot the measured $\delta_G$ versus the transition width $d_{out}$  along with the predicted line for $\delta_{G}$  as expressed in Equation (\ref{eq:delta1}). We clearly see that our prediction works remarkably well for the cases where $\delta_{G}<d_{in}$. 

In the opposite limit, when $\delta_{en}\gg d_{in}$,
\begin{equation}
E_{{\rm CZ}}\simeq\delta_{en}\frac{\text{Ra}_o}{\text{Pr}}Sd_{in}\left(\dfrac{\delta_{en}}{d_{in}}\right),
\end{equation} 
leading to
\begin{equation}
\label{eq:delta2}
\delta_{en}\simeq \left(\dfrac{E_{\rm CZ}\rm{Pr}}{S{\rm Ra}_o}\right)^{1/2}\Rightarrow \delta_G\approx 1.2\left(\dfrac{E_{\rm CZ}\rm{Pr}}{S{\rm Ra}_o}\right)^{1/2}.
\end{equation}
In this limit the downflows penetrate down to the region where $\beta(r) \simeq S$, so it is not surprising to see that $\delta_G$ depends on $S$, but is independent of $d_{out}$. 
Figure \ref{fig:Fig_11}(b) shows the measured $\delta_G$ against the stiffness parameter $S$ along with the scaling given in Equation (\ref{eq:delta2}). We find that  the scaling law $S^{-1/2}$ works for the  simulations in which $\delta_{G}>d_{in}$, but is off by a constant factor. This is not too surprising since the expansion used to obtain Equation (\ref{eq:delta2}) is technically valid only in the limit $(\delta_{en}/d_{in})\rightarrow\infty$, which does not hold true for any of our simulations where $\delta_G$ is fairly close to $d_{in}$. 

Ultimately, we see that  $\delta_G$ is either proportional to $S^{-1/3}$ or to $S^{-1/2}$, implying that it decreases with increasing $S$ in both limits. This is in agreement with the naive expectation that turbulent fluid motions generated in the CZ have a harder time penetrating deeply into a more strongly stratified RZ. These scalings are quite different from the ones proposed by \citet{Zahn91} and \citet{Hurlburt94}, which both argue for a penetration depth (i.e. the depth of the adiabatically stratified layer) scaling as $S^{-1}$, and an overshoot depth (the depth of their thermal adjustment layer) scaling as $S^{-1/4}$. The difference between their theory and our results is relatively easy to understand, however. To start with, their model setup is quite different from ours, relying on changes in the thermal conductivity to drive the transition from a radiative to a convective environment whereas we produce this transition by effectively adding a heating source (see Section \ref{sec:model}). Since their theoretical predictions fundamentally rely on the changes in thermal conductivity, it is not surprising that they would be at odds with our own scalings.  Furthermore, their $S^{-1}$ scaling relies on the existence of an adiabatic penetrative layer, and their $S^{-1/4}$ scaling relies on an exponentially damped overshoot.  Neither of these dynamics are seen here.  Note also that \citet{Rogers2005} presented the results of 2D anelastic simulations of penetrative and overshooting convection, where they confirmed the $S^{-1}$ scaling in the penetrative limit, but report on a much shallower scaling law $\sim S^{-0.04}$ in the moderate- and high-$S$ non-penetrative limit. While their definition of $S$ differs somewhat from that of \citet{Hurlburt94}, that difference cannot fully explain the rather large discrepancy in observed scaling with $S$. Instead, clues to the possible origin of this discrepancy might lie in the applied thermal boundary conditions: \citet{Rogers2005} use isothermal boundary conditions, and state that {\it ``In simulations in which a constant heat flux boundary condition is used at the top, the scaling relation at moderate $S$ values is not as shallow".}  Our findings  then do not  contradict any of these results.

Finally, we note that $\delta_G$ counter-intuitively decreases with increasing Ra$_o$ in both of these limits. Indeed, one would  expect that the increase in the turbulent convective velocities associated with a higher Ra$_o$ would lead to deeper overshooting into the RZ. However, the background stratification of the deep RZ in our model setup scales like $\bar N^2 \simeq S {\rm Ra}_o/\rm{Pr}$ which increases with increasing Ra$_o$ for fixed values of $S$. We therefore see that this second effect dominates the system dynamics, leading to a shallower -- not deeper --  $\delta_G$ as Ra$_o$ increases.

\section{THERMAL MIXING IN THE RZ}
\label{sec:thermal}

\begin{figure}
\centering
\includegraphics[scale=0.25]{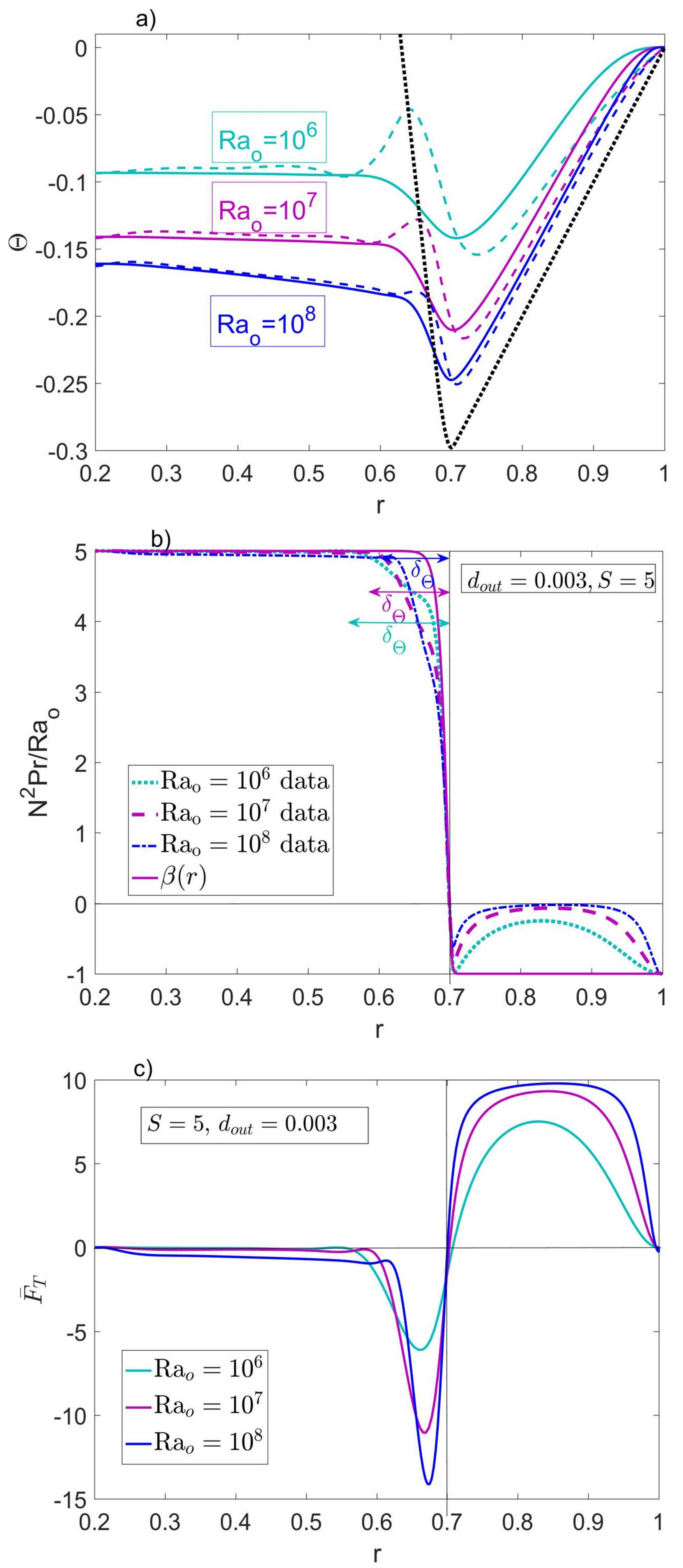}
\caption{a) Plot of the temperature perturbations (where the solid lines correspond to $\bar{\Theta}$ and the dashed lines correspond to $\bar{\Theta}_{down}$) along with the respective adiabatic one (dotted black line),  b) plot of $\bar{N}^2(r){\rm Pr}/{\rm Ra}_o$ along with the respective $N_{\rm{rad}}^2(r){\rm Pr}/{\rm Ra}_o$, and c) plot of the fluxes $\bar{F}_T$,  for $S=5$, $d_{out}=0.003$ and three different Ra$_o$.}
\label{fig:Fig_12}
\end{figure}

In this Section, we focus on quantifying the properties and dependence on input parameters of the regime of partial thermal mixing in the RZ.  Figure \ref{fig:Fig_12}a shows $\bar{\Theta}_{down}$,  and $\bar{\Theta}$, as defined earlier, for the simulation with $S=5$, $d_{out}=0.003$ and  Ra$_o=10^7$ (Case 11, Table \ref{tab:data}) analyzed in Section \ref{sec:samplesim}, along with a more laminar case of Ra$_o=10^6$ (Case 2, Table \ref{tab:data}) and a more turbulent case of Ra$_o=10^8$ (Case 19, Table \ref{tab:data}). Figure \ref{fig:Fig_12}b shows the corresponding buoyancy frequency profiles, and Figure \ref{fig:Fig_12}c shows the associated turbulent temperature flux (see below for its definition and discussion). 

Within the CZ, we find that $\bar N^2$ is closer to 0 (and correspondingly that $\bar{\Theta}$ follows $\Theta_{\rm{ad}}$ more closely) for larger Ra$_o$. This is to be expected since a more turbulent convection zone is more efficient in driving the mean temperature toward an adiabatic state. Meanwhile in the radiative region, we recover the same overall behavior for $\bar N^2$, $\bar{\Theta}_{down}$,  and $\bar{\Theta}$ 
that was already observed in the reference simulation: mixing is not strong enough to cause an extension of the convection zone, but does smooth out the mean stratification down to a depth $\sim \delta_{\Theta}$ below the base of the CZ. We also see that $\delta_{\Theta}$ decreases with increasing Ra$_o$ (and same is true for $\delta_u$), as shown in Table \ref{tab:data}. This trend mirrors the corresponding decrease in $\delta_G$ with increasing Rayleigh number discussed in Section \ref{sec:kinprof}, which was attributed to the increasing stratification of the RZ. Since $\delta_\Theta$ continues to be a good proxy for the depth of the partially thermally mixed region (see Figure \ref{fig:Fig_12}b), our findings therefore imply that the latter becomes shallower with increasing Ra$_o$ .

\begin{figure*}
\centering
\includegraphics[scale=0.1]{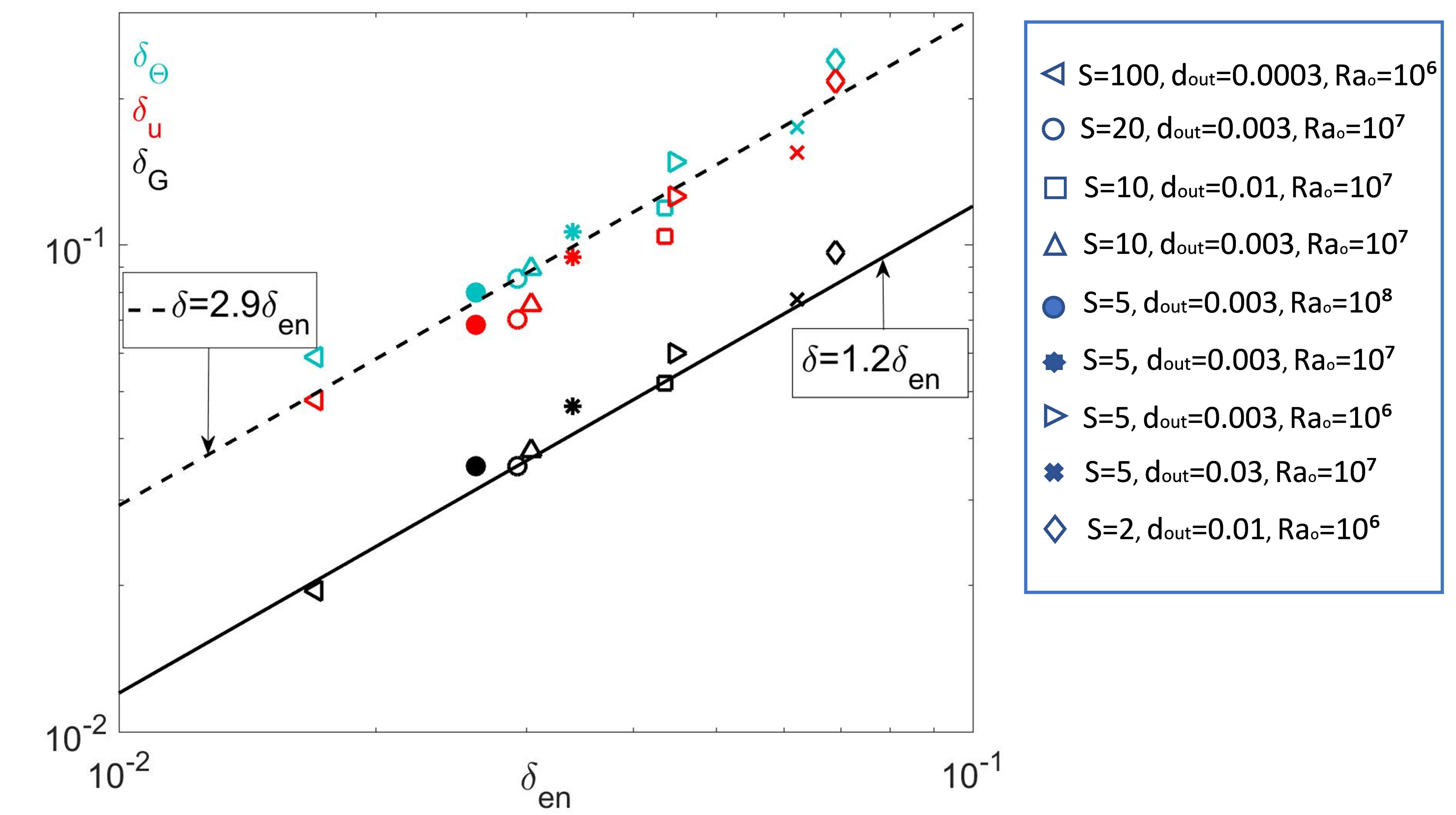}
\caption{Comparison of  $\delta_u$, $\delta_\Theta$ and $\delta_G$ against the estimated $\delta_{en}$, for the simulations indicated on  the legend. Also shown are the best fit to the data, namely $1.2 \delta_{en}$ for $\delta_G$ and $2.9 \delta_{en}$ for $\delta_u$ and $\delta_\Theta$.}
\label{fig:Fig_13}
\end{figure*}

More generally, we have found that $\delta_u$, $\delta_\Theta$ and $\delta_G$ are all very closely related to one another across all of our simulations, and can easily be predicted from the energy-based lengthscale $\delta_{en}$ proposed in Section \ref{sec:kinprof}. Indeed, as shown in Figure \ref{fig:Fig_13}, we  find that $\delta_u \simeq \delta_\Theta \simeq 2.9 \delta_{en}$. In other words, the energy-based argument proposed in 
Section \ref{sec:kinprof} applies equally well to predict the neutral buoyancy point and the stopping depth of individual (strong) downflows, albeit with a somewhat larger prefactor. This provides a very simple way of estimating the depth of the partially thermally-mixed region below the base of the convection zone simply from knowledge of the model parameters. 

A complete model for thermal mixing by convective overshoot requires a quantitative understanding of the strength of such mixing, i.e. of the turbulent heat flux. In this particular model setup, the turbulent heat flux can easily be measured once the system is in a statistically stationary state. Indeed, taking the horizontal average of the thermal energy equation (\ref{eq:heq}) in that state, integrating it once and applying the boundary condition at $r_i$,  we find that 
\begin{equation}
\bar F_T \equiv \overline{u_r \Theta} =\frac{1}{\rm{Pr}} \frac{\partial \bar \Theta}{\partial r} ,
\end{equation}
or in other words, that the sum of the turbulent and diffusive heat fluxes associated with the temperature perturbation $\bar \Theta$ must be zero. This is consistent with our assumption that the total flux through the system is fixed. We then have
\begin{equation}
\bar F_T = \frac{1}{{\rm Pr} } \left( \frac{\bar N^2 {\rm Pr}}{{\rm Ra}_o} - \beta(r) \right), 
\end{equation}
so the turbulent temperature flux $\bar F_T$ can easily be visualized on Figure \ref{fig:Fig_12}b as the (signed) difference between the dashed line and the solid line (times ${\rm Pr}^{-1}$). It is shown, for better clarity, in Figure \ref{fig:Fig_12}c for the same runs. 

\begin{figure*}
\centering
\includegraphics[scale=0.2]{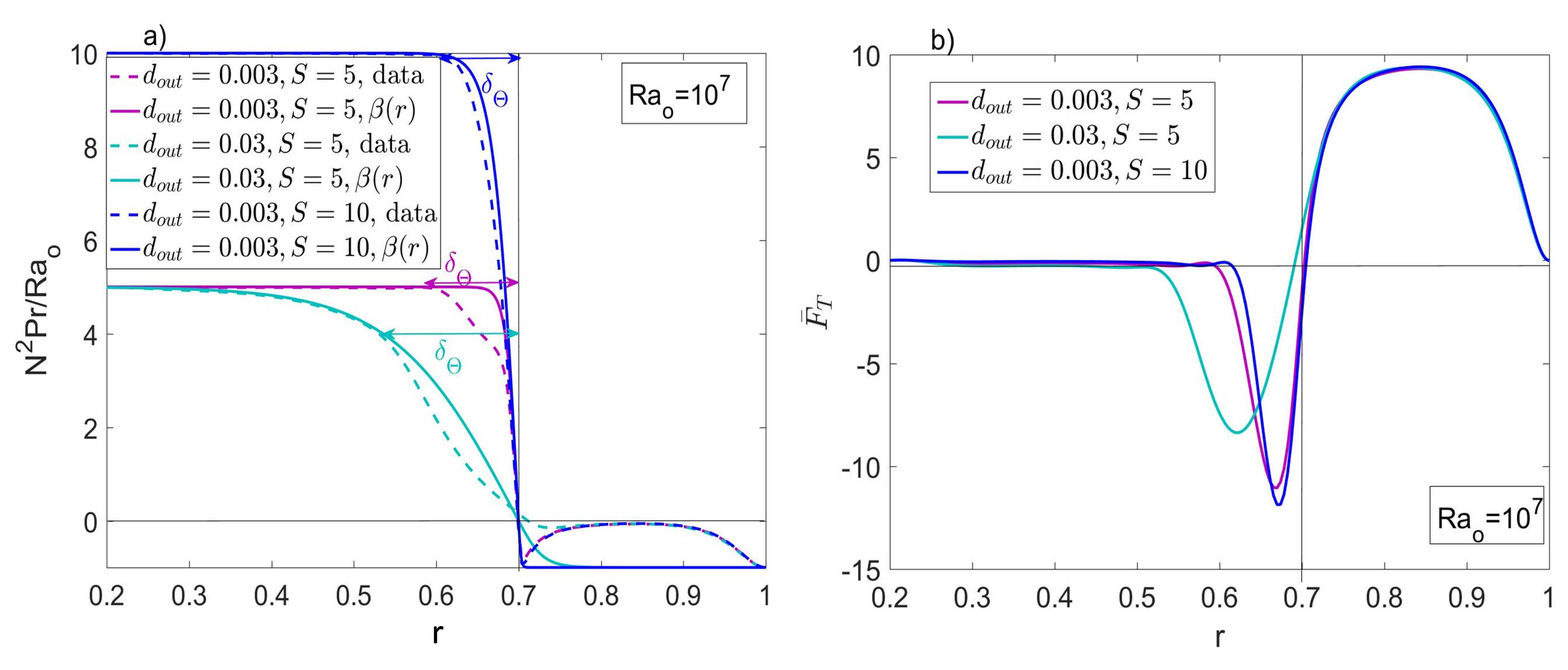}
\caption{a) Comparison of $\bar{N}^2(r){\rm Pr}/{\rm Ra}_o$ with the corresponding background, $\beta(r)$, for simulations with Ra$_o=10^7$, two different values of $S$ (5 and 10), and two different values of  $d_{out}$ (0.003 and 0.03).  b) Corresponding turbulent temperature fluxes for the same simulations.} 
\label{fig:Fig_14}
\end{figure*}

As expected, the temperature flux is generally negative in the radiative zone and positive in the convection zone. It almost always changes sign very close to the radius where $\bar N^2$ changes sign.
In none of the simulations do we see the formation of an {\it extended} stably stratified region subject to substantial positive non-local convective fluxes of the kind reported by \citet{Kapyla}, who called such a layer a ``Deardorff layer" following \citet{Deardorff1966} and \citet{Brandenburg2016}. This difference between our simulations and theirs is probably due to two complementary effects. \citet{Kapyla} ran fully compressible simulations which more realistically capture the asymmetry between weak warm upflows and strong cold downflows than our Boussinesq setup. This asymmetry promotes non-local heat transport by the plumes, allowing the strongest cold downflows to penetrate more coherently and more deeply into the RZ than they would otherwise before warming up. Compressibility is however not a sufficient condition for the formation of a significant Deardorff layer, since none were seen in the compressible simulations of \citet{Brummell} or \citet{Pratt17}.  \citet{Kapyla} explain this, showing that the Deardorff layer is almost absent if the thermal diffusivity profile (or equivalently, the background radiative temperature profile) is fixed and varies abruptly with depth, which is indeed the case in the simulations of \citet{Brummell}. In our numerical setup, which uses the Boussinesq approximation, the asymmetry between upflows and downflows is weak, induced only by the spherical geometry and the boundary conditions. In addition, most of our simulations were run with a transition steepness set by taking $d_{out} = 0.003$, which is very sharp (e.g. see Figure \ref{fig:Fig_12}). We therefore should not expect to see the formation of a Deardorff layer in these cases. We can however detect the existence of one in the largest $d_{out}$ runs (i.e. when $d_{out} = 0.03$; see Figure \ref{fig:Fig_14}) but it remains very shallow. As such, our simulations cannot really probe the dynamics of the Deardorff layer even though we might expect that one should be present in the Sun. 

The magnitude of the turbulent flux below the CZ increases with Ra$_o$, as seen in Figure \ref{fig:Fig_12}c, even though the depth of the mixed layer concurrently decreases. This is not surprising since the r.m.s. velocity of the downflows increases with Ra$_o$ (see  Equation (\ref{eq:ECZmod})). However, the increase of $\bar F_T$ with Rayleigh number is not particularly pronounced, perhaps scaling as ${\bar F_T} \sim {\rm{Ra}}_o^{0.18}$. Within the scope of the simulations shown here, we see that increasing ${\rm Ra}_o$ by a factor of 100 only increases the peak value of $|\bar F_T|$ by a factor of about 2.2 in the RZ. This shows that the turbulent flux itself does not scale as steeply as the r.m.s. velocity (which would lead to $\bar F_T \sim {\rm Ra}_o^{0.36}$), implying in turn that the amplitude of the temperature fluctuations must decrease with increasing Ra$_o$. This can in fact easily be verified in Figure \ref{fig:Fig_12}a, which shows that the profiles of $\bar \Theta$, and $\bar \Theta_{down}$  are much closer to one another at Ra$_o = 10^8$ than at Ra$_o = 10^6$. This result can be explained by noting that turbulence plays an increasingly dominant role at larger Rayleigh number, and has a tendency to homogenize the temperature between upflows and downflows. 
Given the weak dependence on ${\rm Ra}_o$, the range of available simulations is unfortunately not large enough to extract a reliable scaling law between $\bar F_T$ and ${\rm Ra}_o$ -- the latter could be a power law (in which case the power would be of the order of 0.18, as mentioned earlier), but could just as well be logarithmic, or take some other form. As a result, we defer any prediction on the scaling of  $\bar F_T$ with ${\rm Ra}_o$ to future work. Nevertheless, our results point to the crucial importance of accounting for the turbulent mixing between upflows and downflows when modeling mixing by overshooting convection, something that had rarely been taken into account in previous plume models of overshoot \citep{ShavivSalpeter1973,Schmitt84} until the work of \citet{Rempel2004}.

Finally, we explore the dependence of thermal mixing on $S$ and $d_{out}$ in Figure \ref{fig:Fig_14}a, which shows $\bar{N}^2(r){\rm Pr}/{\rm Ra}_o $ and $N_{\rm rad}^2(r){\rm Pr}/{\rm Ra}_o = \beta(r) $ for our typical simulation of $S=5$, $d_{out}=0.003$ and Ra$_o=10^7$ (Case 11, Table \ref{tab:data}) along with one from a simulation with the same $S = 5$ and Ra$_o$ but a larger $d_{out}=0.03$ (shallower transition) (Case 16, Table \ref{tab:data}), and one with the same $d_{out}=0.003$ and Ra$_o$, but a larger $S=10$ (stiffer case) (Case 12, Table \ref{tab:data}). Figure \ref{fig:Fig_14}b shows the corresponding turbulent fluxes for the same simulations. We see that increasing $S$ at fixed $d_{out}$ varies $\delta_\Theta$ a little (so the partially mixed layer below the CZ is somewhat shallower), but the magnitude of the turbulent flux is hardly affected. Increasing $d_{out}$ at fixed $S$ on the other hand has a much larger effect on $\delta_\Theta$ (which increases significantly), and on the fluxes (which decrease by about 25 percent). This shows the importance of smooth versus abrupt transitions in $\beta(r)$, but we have not yet been able to construct a quantitative model to explain these results. 

\section{Summary and discussion}
\label{sec:ccl}

\subsection{Summary}

In this paper, we have presented a series of numerical experiments designed to quantify the interaction between a convective zone and an underlying stably stratified zone, in a spherical geometry and within the context of the Boussinesq approximation. In order to mimic the stellar case, we have used a fixed-flux inner boundary condition at a radius located somewhat above the nuclear burning region, and a fixed-temperature outer boundary condition. For simplicity, all the diffusivities as well as gravity are held constant in the domain, and so is the adiabatic temperature gradient. As a result, a heating source must be invoked in the vicinity of the radiative-convective interface to ensure that the lower part of the domain is indeed stably stratified, while the upper part of the domain is convectively unstable. The selected radial distribution and amplitude of the heating source sets the radiative temperature gradients in the radiative and convective zones respectively, and can be adjusted to create  stable and unstable regions with varying relative stability (quantified through the non-dimensional stiffness parameter $S$), as well as steeper or  shallower transitions between the two (quantified through the non-dimensional transition width $d_{out}$): see Section \ref{sec:model}.  For simplicity, the overall geometry of the system was fixed to mimic the solar case (with the radiative--convective interface located at $r_t = 0.7 r_o$), and we also fixed the Prandtl number ${\rm Pr} = \nu/\kappa = 0.1$ in all of our simulations. The parameters varied were $S$ and $d_{out}$, as well as the global Rayleigh number ${\rm Ra}_o$ (defined in Equation (\ref{eq:Rao})). Increasing ${\rm Ra}_o$ is therefore equivalent to reducing the viscosity and thermal diffusivity concurrently. We explored simulations with ${\rm Ra}_o$ ranging from $10^6$ to $10^8$. Note for comparison that ${\rm Pr} \sim 10^{-6}$ and ${\rm Ra}_o \gg 10^{20}$ in the Sun, so none of the simulations should be used to {\it directly} infer properties of the overshooting convective motions. Instead, we merely seek to understand how the properties of the radiative--convective interface {\it scale} with input parameters, to later attempt an extrapolation of the results to the solar case (while always maintaining some degree of healthy skepticism).

Our simulations all share the same characteristics. We found as in \citet{Korre} that  at fixed aspect ratio and fixed Pr, the mean kinetic energy in the CZ, called $E_{\rm CZ}$, scales as ${\rm Ra}_b^{0.72}$ (see Equation (\ref{eq:ECZmod})), where ${\rm Ra}_b$ is the volume-averaged Rayleigh number within the CZ (see Equation (\ref{eq:Rabdef})), which in this work is quite close to ${\rm Ra}_o$. The total kinetic energy of fluid motions decays below the radiative--convective interface as a Gaussian function of the distance to $r_t$ (see Equation (\ref{eq:Gaussian})) whose width $\delta_G$ can be predicted from first principles using a simple energy argument (aside from a constant of order unity). Indeed, assuming that an average downflow travels a distance $\delta_{en}$ adiabatically from the base of the convection zone until its potential energy equals its estimated initial kinetic energy $E_{\rm CZ}$, we can compute $\delta_{en}$ by solving Equation (\ref{eq:deltaen}). We then showed that, for all available simulations, $\delta_G \simeq 1.2 \delta_{en}$. Through this equation, we can then quantify how $\delta_G$ varies with both the stiffness and steepness of the background stratification profile as well as with the input Rayleigh number. 

We also looked more specifically at how far the {\it strongest} downflows penetrate into {{red}the} RZ, by computing the correlation function $C(\delta)$ between the radial velocity at $r_t$ and a distance $\delta$ away from it. We found that these strong downflows stop at a distance $\delta_u \simeq 2.9 \delta_{en}$ from the base of the convection zone, for any $S$, $d_{out}$ and ${\rm Ra}_o$. This distance $\delta_u$, computed as the first zero of $C(\delta)$, also turns out to correspond to the level of neutral buoyancy for the downflows $\delta_{\Theta}$. The strict correlation discovered between $\delta_u$, $\delta_G$ and $\delta_{en}$ therefore strongly suggests that the simple energetic argument put forward is sufficient to characterize the dynamics of the overshooting plumes. 

We found that the region between $r_t - \delta_u$ and $r_t$ is partially thermally mixed (at these values of the Rayleigh number), resulting in an adjusted buoyancy frequency profile substantially smoother than that of the imposed background. However, we did not see any actual penetration in the traditional definition of the extension of the CZ into the RZ \citep[e.g.][]{Zahn91,Hurlburt94}. This is because the turbulent temperature flux $\bar F_T$ induced by overshooting motions in the RZ remains moderate in all the simulations. We found that it is independent of $S$, and only scales weakly with Rayleigh number (increasing by a factor of about 2 when ${\rm Ra}_o$ increases by a factor of 100), suggesting either a very weak power law ($\bar F_T \propto {\rm Ra}_o^{0.18}$) or a logarithmic dependence on ${\rm Ra}_o$.   

Finally, below $r_t - \delta_u$ the nature of the system dynamics clearly change. The turbulent temperature flux becomes negligible, and the kinetic energy profile is no longer Gaussian, but appears closer to exponential. While weak fluid motions are present, they appear to be more related to the ``damped tail" of linearly unstable convective modes \citep[in the sense described by][for instance]{Freytag96} rather than to internal gravity waves.

\subsection{Comparison with previous numerical experiments} 

As discussed in Section \ref{sec:intro}, there have been quite a few numerical investigations of the dynamics of overshooting and penetrative convection to date. In what follows, we focus on the ones that address the question of overshoot under a convective zone (sometimes referred to as ``undershoot", although we prefer not to use that terminology), rather than above it. These include (among others) the 2D fully compressible simulations of \citet{Hurlburt86}, \citet{Hurlburt94}, \citet{Freytag96} and \citet{Pratt17}, the 3D fully compressible simulations of \citet{Brummell}, \citet{Singh95} \citep[see also][]{Singh98,Saikia2000}, \citet{Kapyla}, the 2D anelastic simulations of \citet{Rogers2005} \citep[see also][]{Rogers2006}, and the 3D ones of \citet{Brun2017}.

Several general conclusions can be drawn from comparing the outcome of these simulations to one another, and to ours. First and foremost is that penetrative convection in the strict definition of the term (i.e. the extension of the convection zone substantially beyond the threshold for linear instability) had so far not been observed in fully turbulent 3D simulations \citep{Brummell,Kapyla,Brun2017}, and this continues to be the case here. As reviewed in Section \ref{sec:intro}, the fact that penetration is seen in 2D at sufficiently low values of $S$ (e.g.  \citet{Rogers2005}) and in very laminar 3D simulations (e.g. \citet{Saikia2000}) can be attributed to the artificially large geometric filling factor of 2D plumes vs. 3D plumes  \citep{Brummell,Rempel2004}. However, none of the existing 3D simulations (including ours) reach particularly high values of the Rayleigh number. Hence, whether this result will continue to hold when progress in supercomputing allows us to simulate convection at much higher Rayleigh numbers remains to be determined (see below for more on this point). 

A second common point between (almost) all simulations is that the kinetic energy of vertical motions within the downflows drops substantially within the CZ as they approach the RZ from above, owing to their lateral deflection, even in low stiffness cases. As a result, the dominant contribution to the total kinetic energy within the RZ is from horizontal flows.  While this may superficially seem at odds with the standard mental picture one may have of overshooting plumes, note that most of the vertical transport is still carried out by the strongest, most-concentrated downflowing motions, as was described in other simulations, e.g. \citet{Brummell,Pratt17}, but the content of these strongest plumes (heat, chemical species) is then advected (and mixed) laterally by turbulent horizontal flows. Precisely how strong these concentrated downflows can get (for given Rayleigh and Prandtl numbers in the CZ) depends on the dimensionality of the simulations and on the compressibility of the fluid (Boussinesq vs. anelastic vs. fully compressible). Since the strength and depth of the downflows control other RZ processes, such as the generation of gravity waves or the formation of a Deardorff layer for instance, it is not surprising to see that the latter are strongly model-dependent, present in some simulations, absent in others. 

A third common point between all simulations is that the depth of the turbulent overshooting layer (as measured by looking at either the kinetic energy profile or the kinetic energy flux below the base of the CZ) does seem to decrease with increasing stiffness $S$, which is an intuitive result. What differs however is the measured scaling law relating this depth to $S$. \citet{Hurlburt94} and \citet{Brummell} both ran direct numerical simulations of overshooting convection in 2D and 3D respectively, where the radiative--convective transition is caused by a sudden change in the thermal conductivity. They both state that their results are consistent with estimates based on a variant of Zahn's theory \citep{Zahn91}, which predicts that the overshooting depth should scale as $\sim S^{-1/4}$ when the total flux through the system is fixed. Our findings are at odds with \cite{Zahn91}, who only considered two distinct regimes: a  very high Pe regime and a  very low Pe regime. The high Pe regime is associated with pure penetrative convection while in the low Pe regime, diffusion dominates and leads to a thermal boundary layer associated with overshooting. Instead, the P\'{e}clet numbers estimated in the overshoot region, and  shown in Table \ref{tab:data}  range from about $0.5$ to approximately $10$. Our reported values of Pe$_{ov}$ are clearly neither in the very low Pe regime nor in the very large Pe regime, thus our findings cannot be directly compared to the regimes found in \cite{Zahn91}. Zahn associated overshooting with a diffusively-dominated regime (Pe$<<1$) whereas here we associate overshooting with Pe$<\sim 1$. \citet{Rogers2005} presented 2D anelastic simulations with fixed temperature boundary conditions, where the radiative--convective transition is also caused by a sudden change in the thermal conductivity, and found a much shallower scaling law $\sim S^{-0.04}$. Meanwhile, in our Boussinesq 3D fixed flux direct numerical simulations, where the transition is driven by the existence of a heating source around $r_t$, we find somewhat steeper scaling laws, with $\sim S^{-1/3}$ or $\sim S^{-1/2}$ depending on whether the background radiative temperature gradient is shallower or steeper, respectively. We believe that the observed difference in the scaling laws reported in these various papers is more likely to be due to the differences in boundary conditions or model setup used rather than compressibility, but this should be verified in future work. It would be interesting for instance to run a comparative study of overshooting and penetrative convection in various systems that all have the same background profile of $N^2$, but that are driven in different ways (i.e. by varying the diffusivities, or the equation of state, or using a heating function, for instance). 

In any case, gaining a better understanding of the scaling of the overshoot depth with $S$ is arguably less important than constraining its scaling with the Rayleigh number, since $S$ is not expected to be too large in stars.  ${\rm Ra}_o$ on the other hand needs to be increased by more than 10 orders of magnitude to reach the stellar regime. Not many studies have systematically looked into this problem. The work of \citet{Brummell} seems to suggest (see their Section 3.7) the approximate scaling $\delta \sim {\rm Ra}^{-0.25}$.
In \citet{Rogers2005}, the situation is  complicated by the fact that the measured scalings with Rayleigh number appear to depend sensitively on $S$: in the less stiff cases, the overshooting layer depth increases with Rayleigh number, but the opposite is true for stiffer cases. The simplicity of our simulations however, easily allows us to vary ${\rm Ra}_o$ independently of all other parameters, and we find that $\delta_G \sim {\rm Ra}_o^{-0.09}$ in the case where the transition is steep, and $\delta_G \sim {\rm Ra}_o^{-0.14}$ when it is shallow. In both cases the overshoot depth therefore decreases with ${\rm Ra}_o$ at fixed $S$ as discussed in Section \ref{sec:kinprof} (see Equations  (\ref{eq:delta1}) and (\ref{eq:delta2}), using $E_{\rm CZ} \sim {\rm Ra}_o^{0.72}$), although the actual power law is quite shallow. 

Finally, we note that very few studies, to our knowledge, really looked into the actual spatial variation of the kinetic energy profile with depth (which is a good proxy for the variation of the mixing coefficient with depth, see  below). \citet{Freytag96} were the first to clearly state that their simulations show an exponential decay of the r.m.s. velocities with depth below the convection zone. They showed that this profile is consistent with these velocities being the stable exponentially-decaying tail of the linearly unstable convective modes. Unfortunately, this also demonstrates that their simulations cannot be in the turbulent regime, a notion that is consistent with a simple visual inspection of their Figures 2-5. By contrast, our simulations are quite turbulent down to about $r_t - \delta_u$. We find that the kinetic energy profile is Gaussian instead of exponential in that region, and only becomes exponential once the fluid motions are sufficiently slow for all nonlinearities to be negligible. 

\subsection{A prescription for mixing by overshoot}

Our numerical results have led us to suggest a very simple Gaussian model for the kinetic energy profile below the base of the convection zone, given by
\begin{equation}
E(r)=E_{\rm CZ} \exp\left(-\frac{(r-r_t)^2}{2\delta_G^2}\right),
\label{eq:Ekindiff}
\end{equation}
where $E_{\rm CZ}$ is the typical kinetic energy of fluid motions within the convection zone (i.e. somewhere within the bulk of the zone). This quantity can for instance be determined from mixing length theory in a stellar evolution code, or from Equation (\ref{eq:ECZmod}) in more idealized Boussinesq setups (recalling that the prefactor could depend on the Prandtl number and on the aspect ratio of the convective region). The lengthscale $\delta_G$ on the other hand can be estimated by using the energy-based lengthscale $\delta_{en}$ discussed in Section \ref{sec:kinprof} (see Equation (\ref{eq:deltapr})), with $\delta_G \simeq \delta_{en}$. A factor of unity relating the two is left here unspecified, and may weakly depend on the Prandtl number and on compressibility. Dimensionally speaking, the lengthscale $\delta_{en}$ can be found by solving the equation 
\begin{equation}
\frac{1}{2} v_{\rm CZ}^2 = - \delta_{en}  \int_{r_t - \delta_{en}}^{r_t} \frac{ \bar g }{H_p} \left( \nabla - \nabla_{\rm ad} \right) dr  ,
 \label{eq:deltaG}
\end{equation}
where $v_{\rm CZ}$ is the convective velocity in the bulk of the convection zone, $\bar g$ is the local gravity, $\nabla={\partial \ln T}/{\partial \ln p}$ is the radiative temperature gradient,  $\nabla_{\rm{ad}}=(\partial \ln T/\partial \ln p)_{\rm{ad}}$ is the adiabatic temperature gradient, and $H_p$ is the pressure scaleheight. With this formula, the computation of the depth $\delta_G$ only depends on the local temperature gradient as well as standard variables returned by stellar evolution codes, rather than the manner in which this temperature gradient (and the CZ-RZ transition) is actually generated. 
 Note that this energy-based argument for estimating the overshoot depth is ultimately quite standard; it recovers, for instance, that of \citet{Christensen-Dalsgaard2011} (see their equation 18) if $\nabla-\nabla_{\rm ad}$ is taken to be approximately constant below the base of the convection zone, in which case $\delta_{en}$ satisfies
\begin{equation}
\frac{ \delta_{en}}{H_p}   =  \left( \frac{v_{\rm CZ}^2 }{2 \bar g H_p  |\nabla - \nabla_{\rm ad} |} \right)^{1/2}.
 \label{eq:deltaGCD}
\end{equation}
If, on the other hand, $\nabla - \nabla_{\rm ad}$ is assumed to vary linearly with depth below the CZ, with $\nabla - \nabla_{\rm ad} \simeq  \eta (r - r_t)$, then 
\begin{equation}
\frac{\delta_{en}  }{H_p}   = \left(  \frac{v_{\rm CZ}^2}{\bar g H_p^2 \eta }  \right)^{1/3}.
 \label{eq:deltaGnew}
\end{equation}
From the kinetic energy profile (\ref{eq:Ekindiff}), we can then form  a diffusion coefficient to model compositional mixing by overshooting motions
\begin{equation}
D_{ov}(r) = D_{\rm CZ} \exp\left(-\frac{(r-r_t)^2}{2\delta_G^2}\right) ,
\end{equation} 
assuming that $D_{ov} \propto v_{\rm CZ}^2 \tau_{\rm CZ}$ as in \citet{Freytag96}, where $\tau_{\rm CZ}$ is some convective turnover timescale just above the base of the convection zone. 

In order to apply Equation (\ref{eq:deltaGnew}) to the Sun, we extract all the relevant quantities from the interface between the interior radiation zone and convective envelope of a 1 solar mass Main-Sequence model computed with MESA\footnote{Version 6794.} \citep{Paxton2011,Paxton2013}. We find that $v_{\rm CZ} \simeq 6,000$cm/s in the bulk of the convection zone, and $g \simeq 50,000$cm/s$^{2}$, $H_p \simeq 5 \times 10^9$cm, and $\eta \simeq  10^{-10}$cm$^{-1}$ near the interface, leading to $\delta_{en}/H_p \simeq 0.006$. Similar calculations made at the interface between the interior convective zone and radiative envelope of a 2 solar mass Main-Sequence model yield $v_{\rm CZ} \simeq 7,000$cm/s, $g \simeq 200,000$cm/s$^{2}$, $H_p \simeq 5 \times 10^9$cm, and $\eta \simeq  10^{-11}$cm$^{-1}$, leading to $\delta_{en}/H_p \simeq 0.01$. In both cases, $\delta_{en}$ (and by definition $\delta_G$) is quite a small fraction of a pressure scaleheight, and would result in much shallower predictions for the depth of the overshoot-mixed layer than what is commonly assumed in stellar evolution models (e.g. from the model of \citet{Herwig2000} with $f \simeq 0.1 H_p$). Even shallower predictions would be obtained using values of $v_{\rm CZ}$ taken closer to the edge of the convective region. Whether overshoot is in fact as shallow as predicted in real stars remains to be determined. As discussed in Section \ref{sec:intro}, it is not unlikely that moving beyond the Boussinesq approximation could result in a somewhat larger overshoot depth than what we currently see in the simulations, simply because of the pressure-induced enhancement of the asymmetry between narrow downflows and broad upflows. In addition, since $\delta_{en}$ depends sensitively on $v_{\rm CZ}$, the reliability of our model predictions effectively depends on the reliability of mixing-length theory \citep{BohmVitense1958,CoxGiuli1968} in estimating the typical velocities of convective motions deep within  a star. Asteroseismology will hopefully help constrain the latter in the coming years. Nevertheless, it is difficult to see how the overshoot depth could vary substantially away from $\delta_{en}$ predicted using a simple energy balance argument. 
It is worth remembering at this point that Zahn's original models \citep{Zahn91,Hurlburt94} also predict a very shallow overshoot layer (in the strict definition of the term) -- but that layer only starts beyond a thermally-mixed penetration layer which can be much larger (at least for the smaller values of $S$). As such, our findings (in terms of strict overshoot) are not inconsistent with observations of substantial mixing beyond the edge of a convective region \citep{Liu14,Deheuvels2016}, as long as these observations are interpreted as evidence for penetration (rather than overshoot).

 As discussed in Section \ref{sec:thermal}, estimating the amount of {\it thermal} mixing below the CZ (and therefore quantifying penetration) is much more complicated, as this requires knowledge not only of the velocities, but also of the typical temperature fluctuations associated with upflows and downflows relative to the background profile, which in turn depend on the relative importance of both small-scale horizontal turbulent mixing and thermal diffusion, as well as the global thermal equilibrium. This cannot be done using simple local energetic/thermal balance arguments, and it seems that the only way forward  is to analyze the results of numerical simulations to create an empirical model for the heat flux. The problem with this approach, however, is that it is very sensitive to the model setup used (i.e. compressible vs. anelastic vs. Boussinesq, 2D vs. 3D, boundary conditions, method for generating the CZ-RZ transition), as noted by the rather vast discrepancies in results obtained in the numerical experiments discussed in Section \ref{sec:intro}. Further work will be required to better understand the causes of this sensitivity, and to determine what results can and cannot be carried over (qualitatively and quantitatively) from idealized models to more realistic stellar environments. 
 
 {\it Within the scope of numerical simulations run in the same setup as ours,} we could tentatively extrapolate our results to estimate the magnitude of the turbulent temperature flux $\bar F_T$ induced below $r_t$ by the convective motions. However, we found that the latter only varies very weakly with Rayleigh number, to the extent that we are unable to propose any definite model for the former. If a power law is assumed, then our results suggest that $\bar F_T \propto {\rm Ra}_o^{0.18}$. If that scaling holds, we predict that it may be possible to see convective penetration in Boussinesq convection at higher Rayleigh numbers (holding the Prandtl number constant). Indeed, taking our reference simulation (Pr = 0.1, $S = 5$, $d_{out} = 0.003$, Ra$_o = 10^7$) for instance, we see that the turbulent flux would have to be about 5 times larger than it is to drive the profile of $\bar N^2$ towards an adiabat below the base of the CZ, which would require an input Rayleigh number (defined as in Equation (\ref{eq:Rao})) of the order of about $10^{11}$. Another way of looking at the problem is to estimate how the P\'{e}clet number varies with Ra$_o$. Given that Pe$_{ov}=u_{rms}\delta_G$Pr, where  $u_{rms}\propto$Ra$_b^{0.36}$ (see Eq. (\ref{eq:ECZmod})) and $\delta_G\approx \delta_{en}$ (where $\delta_{en}$ is given by Eq. (\ref{eq:delta1}), for instance), we find that Pe$_{ov}\propto$Ra$_o^{0.27}$. It would require a  Rayleigh number $10^4$ times larger, therefore  a P\'{e}clet number $10$ times larger than what we currently have in order to get a fully mixed region, i.e. pure penetration.  This is quite large, but may actually be achievable in the not-too-distant future\footnote{Recall that Ra$_o$ is based on the lengthscale $r_o$ rather than the width of the CZ, so the effective Rayleigh number of our simulations is smaller than Ra$_o$.} (especially if one were to use a reduced computational domain consisting of a wedge, rather than a full sphere).

\bibliographystyle{mnras}

\bibliography{biblio}

\providecommand{\noopsort}[1]{}\providecommand{\singleletter}[1]{#1}%
\begin{thebibliography}{}
\makeatletter
\relax
\def\mn@urlcharsother{\let\do\@makeother \do\$\do\&\do\#\do\^\do\_\do\%\do\~}
\def\mn@doi{\begingroup\mn@urlcharsother \@ifnextchar [ {\mn@doi@}
  {\mn@doi@[]}}
\def\mn@doi@[#1]#2{\def\@tempa{#1}\ifx\@tempa\@empty \href
  {http://dx.doi.org/#2} {doi:#2}\else \href {http://dx.doi.org/#2} {#1}\fi
  \endgroup}
\def\mn@eprint#1#2{\mn@eprint@#1:#2::\@nil}
\def\mn@eprint@arXiv#1{\href {http://arxiv.org/abs/#1} {{\tt arXiv:#1}}}
\def\mn@eprint@dblp#1{\href {http://dblp.uni-trier.de/rec/bibtex/#1.xml}
  {dblp:#1}}
\def\mn@eprint@#1:#2:#3:#4\@nil{\def\@tempa {#1}\def\@tempb {#2}\def\@tempc
  {#3}\ifx \@tempc \@empty \let \@tempc \@tempb \let \@tempb \@tempa \fi \ifx
  \@tempb \@empty \def\@tempb {arXiv}\fi \@ifundefined
  {mn@eprint@\@tempb}{\@tempb:\@tempc}{\expandafter \expandafter \csname
  mn@eprint@\@tempb\endcsname \expandafter{\@tempc}}}

\bibitem[\protect\citeauthoryear{{Ahrens}, {Stix}  \& {Thorn}}{{Ahrens}
  et~al.}{1992}]{Ahrens92}
{Ahrens} B.,  {Stix} M.,   {Thorn} M.,  1992, \aap, 264, 673

\bibitem[\protect\citeauthoryear{{Aubert}, {Aurnou}  \& {Wicht}}{{Aubert}
  et~al.}{2008}]{Parody1}
{Aubert} J.,  {Aurnou} J.,   {Wicht} J.,  2008, Geophysical Journal
  International, 172, 945

\bibitem[\protect\citeauthoryear{{Baraffe}, {Pratt}, {Goffrey}, {Constantino},
  {Folini}, {Popov}, {Walder}  \& {Viallet}}{{Baraffe}
  et~al.}{2017}]{Baraffe17}
{Baraffe} I.,  {Pratt} J.,  {Goffrey} T.,  {Constantino} T.,  {Folini} D.,
  {Popov} M.~V.,  {Walder} R.,   {Viallet} M.,  2017, \apjl, 845, L6

\bibitem[\protect\citeauthoryear{{B{\"o}hm-Vitense}}{{B{\"o}hm-Vitense}}{1958}]{BohmVitense1958}
{B{\"o}hm-Vitense} E.,  1958, \zap, 46, 108

\bibitem[\protect\citeauthoryear{{Brandenburg}}{{Brandenburg}}{2016}]{Brandenburg2016}
{Brandenburg} A.,  2016, \apj, 832, 6

\bibitem[\protect\citeauthoryear{{Browning}, {Brun}  \& {Toomre}}{{Browning}
  et~al.}{2004}]{Browning2004}
{Browning} M.~K.,  {Brun} A.~S.,   {Toomre} J.,  2004, \apj, 601, 512

\bibitem[\protect\citeauthoryear{{Browning}, {Miesch}, {Brun}  \&
  {Toomre}}{{Browning} et~al.}{2006}]{Browning2006}
{Browning} M.~K.,  {Miesch} M.~S.,  {Brun} A.~S.,   {Toomre} J.,  2006, \apjl,
  648, L157

\bibitem[\protect\citeauthoryear{{Browning}, {Brun}, {Miesch}  \&
  {Toomre}}{{Browning} et~al.}{2007}]{Browning2007}
{Browning} M.~K.,  {Brun} A.~S.,  {Miesch} M.~S.,   {Toomre} J.,  2007,
  Astronomische Nachrichten, 328, 1100

\bibitem[\protect\citeauthoryear{{Brummell}, {Clune}  \& {Toomre}}{{Brummell}
  et~al.}{2002}]{Brummell}
{Brummell} N.~H.,  {Clune} T.~L.,   {Toomre} J.,  2002, \apj, 570, 825

\bibitem[\protect\citeauthoryear{{Brun}, {Browning}  \& {Toomre}}{{Brun}
  et~al.}{2005}]{Brun2005}
{Brun} A.~S.,  {Browning} M.~K.,   {Toomre} J.,  2005, \apj, 629, 461

\bibitem[\protect\citeauthoryear{{Brun}, {Miesch}  \& {Toomre}}{{Brun}
  et~al.}{2011}]{BMT11}
{Brun} A.~S.,  {Miesch} M.~S.,   {Toomre} J.,  2011, \apj, 742, 79

\bibitem[\protect\citeauthoryear{{Brun} et~al.,}{{Brun}
  et~al.}{2017}]{Brun2017}
{Brun} A.~S.,  et~al., 2017, \apj, 836, 192

\bibitem[\protect\citeauthoryear{{Charbonneau} \& {MacGregor}}{{Charbonneau} \&
  {MacGregor}}{1997}]{Charbonneau97}
{Charbonneau} P.,  {MacGregor} K.~B.,  1997, \apj, 486, 502

\bibitem[\protect\citeauthoryear{{Christensen-Dalsgaard}, {Gough}  \&
  {Thompson}}{{Christensen-Dalsgaard} et~al.}{1991}]{Christensen-Dalsgaard91}
{Christensen-Dalsgaard} J.,  {Gough} D.~O.,   {Thompson} M.~J.,  1991, \apj,
  378, 413

\bibitem[\protect\citeauthoryear{{Christensen-Dalsgaard}, {Monteiro}, {Rempel}
  \& {Thompson}}{{Christensen-Dalsgaard}
  et~al.}{2011}]{Christensen-Dalsgaard2011}
{Christensen-Dalsgaard} J.,  {Monteiro} M.~J.~P.~F.~G.,  {Rempel} M.,
  {Thompson} M.~J.,  2011, \mnras, 414, 1158

\bibitem[\protect\citeauthoryear{{Christensen-Dalsgaard}, {Gough}  \&
  {Knudstrup}}{{Christensen-Dalsgaard}
  et~al.}{2018}]{Christensen-Dalsgaard2018}
{Christensen-Dalsgaard} J.,  {Gough} D.~O.,   {Knudstrup} E.,  2018, \mnras,
  477, 3845

\bibitem[\protect\citeauthoryear{{Cogan}}{{Cogan}}{1975}]{Cogan75}
{Cogan} B.~C.,  1975, \apj, 201, 637

\bibitem[\protect\citeauthoryear{{Couston}, {Lecoanet}, {Favier}  \& {Le
  Bars}}{{Couston} et~al.}{2017}]{Couston17}
{Couston} L.-A.,  {Lecoanet} D.,  {Favier} B.,   {Le Bars} M.,  2017, Physical
  Review Fluids, 2, 094804

\bibitem[\protect\citeauthoryear{{Cox} \& {Giuli}}{{Cox} \&
  {Giuli}}{1968}]{CoxGiuli1968}
{Cox} J.~P.,  {Giuli} R.~T.,  1968, {Principles of stellar structure }

\bibitem[\protect\citeauthoryear{{Deardorff}}{{Deardorff}}{1966}]{Deardorff1966}
{Deardorff} J.~W.,  1966, Journal of Atmospheric Sciences, 23, 503

\bibitem[\protect\citeauthoryear{{Deardorff}, {Willis}  \& {Lilly}}{{Deardorff}
  et~al.}{1969}]{Deardorff69}
{Deardorff} J.~W.,  {Willis} G.~E.,   {Lilly} D.~K.,  1969, Journal of Fluid
  Mechanics, 35, 7

\bibitem[\protect\citeauthoryear{{Deheuvels}, {Brand{\~a}o}, {Silva Aguirre},
  {Ballot}, {Michel}, {Cunha}, {Lebreton}  \& {Appourchaux}}{{Deheuvels}
  et~al.}{2016}]{Deheuvels2016}
{Deheuvels} S.,  {Brand{\~a}o} I.,  {Silva Aguirre} V.,  {Ballot} J.,  {Michel}
  E.,  {Cunha} M.~S.,  {Lebreton} Y.,   {Appourchaux} T.,  2016, \aap, 589, A93

\bibitem[\protect\citeauthoryear{{Dormy}, {Cardin}  \& {Jault}}{{Dormy}
  et~al.}{1998}]{Parody2}
{Dormy} E.,  {Cardin} P.,   {Jault} D.,  1998, Earth and Planetary Science
  Letters, 160, 15

\bibitem[\protect\citeauthoryear{{Freytag}, {Ludwig}  \& {Steffen}}{{Freytag}
  et~al.}{1996}]{Freytag96}
{Freytag} B.,  {Ludwig} H.-G.,   {Steffen} M.,  1996, \aap, 313, 497

\bibitem[\protect\citeauthoryear{{Ghizaru}, {Charbonneau}  \&
  {Smolarkiewicz}}{{Ghizaru} et~al.}{2010}]{Ghizaru2010}
{Ghizaru} M.,  {Charbonneau} P.,   {Smolarkiewicz} P.~K.,  2010, \apjl, 715,
  L133

\bibitem[\protect\citeauthoryear{{Gough}, {Spiegel}  \& {Toomre}}{{Gough}
  et~al.}{1975}]{GST75}
{Gough} D.~O.,  {Spiegel} E.~A.,   {Toomre} J.,  1975, Journal of Fluid
  Mechanics, 68, 695

\bibitem[\protect\citeauthoryear{{Herring}}{{Herring}}{1963}]{Herring63}
{Herring} J.~R.,  1963, Journal of Atmospheric Sciences, 20, 325

\bibitem[\protect\citeauthoryear{{Herwig}}{{Herwig}}{2000}]{Herwig2000}
{Herwig} F.,  2000, \aap, 360, 952

\bibitem[\protect\citeauthoryear{{Hurlburt}, {Toomre}  \&
  {Massaguer}}{{Hurlburt} et~al.}{1986}]{Hurlburt86}
{Hurlburt} N.~E.,  {Toomre} J.,   {Massaguer} J.~M.,  1986, \apj, 311, 563

\bibitem[\protect\citeauthoryear{{Hurlburt}, {Toomre}, {Massaguer}  \&
  {Zahn}}{{Hurlburt} et~al.}{1994}]{Hurlburt94}
{Hurlburt} N.~E.,  {Toomre} J.,  {Massaguer} J.~M.,   {Zahn} J.-P.,  1994,
  \apj, 421, 245

\bibitem[\protect\citeauthoryear{{K{\"a}pyl{\"a}}, {Rheinhardt}, {Brandenburg},
  {Arlt}, {K{\"a}pyl{\"a}}, {Lagg}, {Olspert}  \& {Warnecke}}{{K{\"a}pyl{\"a}}
  et~al.}{2017}]{Kapyla}
{K{\"a}pyl{\"a}} P.~J.,  {Rheinhardt} M.,  {Brandenburg} A.,  {Arlt} R.,
  {K{\"a}pyl{\"a}} M.~J.,  {Lagg} A.,  {Olspert} N.,   {Warnecke} J.,  2017,
  \apjl, 845, L23

\bibitem[\protect\citeauthoryear{{Korre}, {Brummell}  \& {Garaud}}{{Korre}
  et~al.}{2017}]{Korre}
{Korre} L.,  {Brummell} N.,   {Garaud} P.,  2017, \pre, 96, 033104

\bibitem[\protect\citeauthoryear{{Latour}, {Spiegel}, {Toomre}  \&
  {Zahn}}{{Latour} et~al.}{1976}]{LSTZ76}
{Latour} J.,  {Spiegel} E.~A.,  {Toomre} J.,   {Zahn} J.-P.,  1976, \apj, 207,
  233

\bibitem[\protect\citeauthoryear{{Latour}, {Toomre}  \& {Zahn}}{{Latour}
  et~al.}{1981}]{LTZ81}
{Latour} J.,  {Toomre} J.,   {Zahn} J.-P.,  1981, \apj, 248, 1081

\bibitem[\protect\citeauthoryear{{Liu} et~al.,}{{Liu} et~al.}{2014}]{Liu14}
{Liu} Z.,  et~al., 2014, \apj, 780, 152

\bibitem[\protect\citeauthoryear{{Maeder}}{{Maeder}}{1975}]{Maeder75}
{Maeder} A.,  1975, \aap, 40, 303

\bibitem[\protect\citeauthoryear{{Malkus}}{{Malkus}}{1960}]{Malkus}
{Malkus} W.~V.~R.,  1960, Aerodynamic Phenomena in Stellar Atmospheres, IAU
  Symp. 12, Bologna, Italy

\bibitem[\protect\citeauthoryear{{Massaguer}, {Latour}, {Toomre}  \&
  {Zahn}}{{Massaguer} et~al.}{1984}]{MLTZ84}
{Massaguer} J.~M.,  {Latour} J.,  {Toomre} J.,   {Zahn} J.-P.,  1984, \aap,
  140, 1

\bibitem[\protect\citeauthoryear{{Miesch}, {Brun}  \& {Toomre}}{{Miesch}
  et~al.}{2006}]{Miesch2006}
{Miesch} M.~S.,  {Brun} A.~S.,   {Toomre} J.,  2006, \apj, 641, 618

\bibitem[\protect\citeauthoryear{{Monteiro} \& {Thompson}}{{Monteiro} \&
  {Thompson}}{1998}]{Monteiro1998}
{Monteiro} M.~J.~P.~F.~G.,  {Thompson} M.~J.,  1998, in {Korzennik} S.,  ed.,
  ESA Special Publication Vol. 418, Structure and Dynamics of the Interior of
  the Sun and Sun-like Stars. p.~819

\bibitem[\protect\citeauthoryear{{Monteiro}, {Christensen-Dalsgaard}  \&
  {Thompson}}{{Monteiro} et~al.}{1994}]{Monteiro1994}
{Monteiro} M.~J.~P.~F.~G.,  {Christensen-Dalsgaard} J.,   {Thompson} M.~J.,
  1994, \aap, 283, 247

\bibitem[\protect\citeauthoryear{{Moore} \& {Weiss}}{{Moore} \&
  {Weiss}}{1973}]{MW73}
{Moore} D.~R.,  {Weiss} N.~O.,  1973, Journal of Fluid Mechanics, 61, 553

\bibitem[\protect\citeauthoryear{{Morton}, {Taylor}  \& {Turner}}{{Morton}
  et~al.}{1956}]{Morton56}
{Morton} B.~R.,  {Taylor} G.,   {Turner} J.~S.,  1956, Proceedings of the Royal
  Society of London Series A, 234, 1

\bibitem[\protect\citeauthoryear{{Musman}}{{Musman}}{1968}]{Musman68}
{Musman} S.,  1968, Journal of Fluid Mechanics, 31, 343

\bibitem[\protect\citeauthoryear{{Muthsam}, {Goeb}, {Kupka}, {Liebich}  \&
  {Zoechling}}{{Muthsam} et~al.}{1995}]{Muthsam95}
{Muthsam} H.~J.,  {Goeb} W.,  {Kupka} F.,  {Liebich} W.,   {Zoechling} J.,
  1995, \aap, 293, 127

\bibitem[\protect\citeauthoryear{{Parker}}{{Parker}}{1993}]{Parker93}
{Parker} E.~N.,  1993, \apj, 408, 707

\bibitem[\protect\citeauthoryear{{Paxton}, {Bildsten}, {Dotter}, {Herwig},
  {Lesaffre}  \& {Timmes}}{{Paxton} et~al.}{2011}]{Paxton2011}
{Paxton} B.,  {Bildsten} L.,  {Dotter} A.,  {Herwig} F.,  {Lesaffre} P.,
  {Timmes} F.,  2011, \apjs, 192, 3

\bibitem[\protect\citeauthoryear{{Paxton} et~al.,}{{Paxton}
  et~al.}{2013}]{Paxton2013}
{Paxton} B.,  et~al., 2013, \apjs, 208, 4

\bibitem[\protect\citeauthoryear{{Pinsonneault}}{{Pinsonneault}}{1997}]{Pinsonneault}
{Pinsonneault} M.,  1997, \araa, 35, 557

\bibitem[\protect\citeauthoryear{{Pratt}, {Baraffe}, {Goffrey}, {Constantino},
  {Viallet}, {Popov}, {Walder}  \& {Folini}}{{Pratt} et~al.}{2017}]{Pratt17}
{Pratt} J.,  {Baraffe} I.,  {Goffrey} T.,  {Constantino} T.,  {Viallet} M.,
  {Popov} M.~V.,  {Walder} R.,   {Folini} D.,  2017, \aap, 604, A125

\bibitem[\protect\citeauthoryear{{Racine}, {Charbonneau}, {Ghizaru}, {Bouchat}
  \& {Smolarkiewicz}}{{Racine} et~al.}{2011}]{Racine2011}
{Racine} {\'E}.,  {Charbonneau} P.,  {Ghizaru} M.,  {Bouchat} A.,
  {Smolarkiewicz} P.~K.,  2011, \apj, 735, 46

\bibitem[\protect\citeauthoryear{{Rempel}}{{Rempel}}{2004}]{Rempel2004}
{Rempel} M.,  2004, \apj, 607, 1046

\bibitem[\protect\citeauthoryear{{Renzini}}{{Renzini}}{1987}]{Renzini}
{Renzini} A.,  1987, \aap, 188, 49

\bibitem[\protect\citeauthoryear{{Roberts}}{{Roberts}}{1966}]{Roberts66}
{Roberts} P.~H.,  1966, in {Donnelly} R.~J.,  {Herman} R.,   {Prigogine} I.,
  eds, Non-Equilibrium Thermodynamics, Variational Techniques, and Stability.
  p.~125

\bibitem[\protect\citeauthoryear{{Rogers} \& {Glatzmaier}}{{Rogers} \&
  {Glatzmaier}}{2005}]{Rogers2005}
{Rogers} T.~M.,  {Glatzmaier} G.~A.,  2005, \apj, 620, 432

\bibitem[\protect\citeauthoryear{{Rogers}, {Glatzmaier}  \& {Jones}}{{Rogers}
  et~al.}{2006}]{Rogers2006}
{Rogers} T.~M.,  {Glatzmaier} G.~A.,   {Jones} C.~A.,  2006, \apj, 653, 765

\bibitem[\protect\citeauthoryear{{Saikia}, {Singh}, {Chan}, {Roxburgh}  \&
  {Srivastava}}{{Saikia} et~al.}{2000}]{Saikia2000}
{Saikia} E.,  {Singh} H.~P.,  {Chan} K.~L.,  {Roxburgh} I.~W.,   {Srivastava}
  M.~P.,  2000, \apj, 529, 402

\bibitem[\protect\citeauthoryear{{Schmitt}, {Rosner}  \& {Bohn}}{{Schmitt}
  et~al.}{1984}]{Schmitt84}
{Schmitt} J.~H.~M.~M.,  {Rosner} R.,   {Bohn} H.~U.,  1984, \apj, 282, 316

\bibitem[\protect\citeauthoryear{{Shaviv} \& {Salpeter}}{{Shaviv} \&
  {Salpeter}}{1973}]{ShavivSalpeter1973}
{Shaviv} G.,  {Salpeter} E.~E.,  1973, \apj, 184, 191

\bibitem[\protect\citeauthoryear{{Silva Aguirre}, {Ballot}, {Serenelli}  \&
  {Weiss}}{{Silva Aguirre} et~al.}{2011}]{SilvaAguirre2011}
{Silva Aguirre} V.,  {Ballot} J.,  {Serenelli} A.~M.,   {Weiss} A.,  2011,
  \aap, 529, A63

\bibitem[\protect\citeauthoryear{{Singh}, {Roxburgh}  \& {Chan}}{{Singh}
  et~al.}{1995}]{Singh95}
{Singh} H.~P.,  {Roxburgh} I.~W.,   {Chan} K.~L.,  1995, \aap, 295, 703

\bibitem[\protect\citeauthoryear{{Singh}, {Roxburgh}  \& {Chan}}{{Singh}
  et~al.}{1998}]{Singh98}
{Singh} H.~P.,  {Roxburgh} I.~W.,   {Chan} K.~L.,  1998, \aap, 340, 178

\bibitem[\protect\citeauthoryear{{Spiegel}}{{Spiegel}}{1963}]{Spiegel63}
{Spiegel} E.~A.,  1963, \apj, 138, 216

\bibitem[\protect\citeauthoryear{{Spiegel} \& {Veronis}}{{Spiegel} \&
  {Veronis}}{1960}]{SV}
{Spiegel} E.~A.,  {Veronis} G.,  1960, \apj, 131, 442

\bibitem[\protect\citeauthoryear{{Spite} \& {Spite}}{{Spite} \&
  {Spite}}{1982}]{Spite82}
{Spite} F.,  {Spite} M.,  1982, \aap, 115, 357

\bibitem[\protect\citeauthoryear{{Straus}, {Blake}  \& {Schramm}}{{Straus}
  et~al.}{1976}]{Straus76}
{Straus} J.~M.,  {Blake} J.~B.,   {Schramm} D.~N.,  1976, \apj, 204, 481

\bibitem[\protect\citeauthoryear{{Strugarek}, {Brun}  \& {Zahn}}{{Strugarek}
  et~al.}{2011}]{Strugarek2011}
{Strugarek} A.,  {Brun} A.~S.,   {Zahn} J.-P.,  2011, \aap, 532, A34

\bibitem[\protect\citeauthoryear{{Sukhbold} \& {Woosley}}{{Sukhbold} \&
  {Woosley}}{2014}]{Sukhbold2014}
{Sukhbold} T.,  {Woosley} S.~E.,  2014, \apj, 783, 10

\bibitem[\protect\citeauthoryear{{Toomre}, {Zahn}, {Latour}  \&
  {Spiegel}}{{Toomre} et~al.}{1976}]{TZLS76}
{Toomre} J.,  {Zahn} J.-P.,  {Latour} J.,   {Spiegel} E.~A.,  1976, \apj, 207,
  545

\bibitem[\protect\citeauthoryear{{Toomre}, {Gough}  \& {Spiegel}}{{Toomre}
  et~al.}{1977}]{TGS77}
{Toomre} J.,  {Gough} D.~O.,   {Spiegel} E.~A.,  1977, Journal of Fluid
  Mechanics, 79, 1

\bibitem[\protect\citeauthoryear{{Toomre}, {Gough}  \& {Spiegel}}{{Toomre}
  et~al.}{1982}]{TGS82}
{Toomre} J.,  {Gough} D.~O.,   {Spiegel} E.~A.,  1982, Journal of Fluid
  Mechanics, 125, 99

\bibitem[\protect\citeauthoryear{{Townsend}}{{Townsend}}{1964}]{Townsend}
{Townsend} A.~A.,  1964, Quarterly Journal of the Royal Meteorological Society,
  90, 248

\bibitem[\protect\citeauthoryear{{Veronis}}{{Veronis}}{1963}]{Veronis63}
{Veronis} G.,  1963, \apj, 137, 641

\bibitem[\protect\citeauthoryear{{Zahn}}{{Zahn}}{1991}]{Zahn91}
{Zahn} J.-P.,  1991, \aap, 252, 179

\bibitem[\protect\citeauthoryear{{Zahn}, {Toomre}  \& {Latour}}{{Zahn}
  et~al.}{1982}]{ZTL82}
{Zahn} J.-P.,  {Toomre} J.,   {Latour} J.,  1982, Geophysical and Astrophysical
  Fluid Dynamics, 22, 159

\bibitem[\protect\citeauthoryear{{van Ballegooijen}}{{van
  Ballegooijen}}{1982}]{vanBallegooijen}
{van Ballegooijen} A.~A.,  1982, \aap, 113, 99

\makeatother
\end{thebibliography}

\section*{Acknowledgements} 
The authors thank C. Guervilly for setting up the two-layered (CZ-RZ) configuration in the PARODY code and for  technical and scientific advice. This work was
financially supported by NASA Grant No. NNX14AG08G. Simulations were run on the Hyades cluster at UCSC, purchased using the NSF  grant No. AST-1229745, as well as on the XSEDE Stampede 2  Texas Advanced Computing Center (TACC) at 
The University of Texas at Austin.

\bsp	
\label{lastpage}
\end{document}